\documentclass[reprint,aps,prd,nofootinbib]{revtex4-2}
\usepackage{graphicx, epsfig}
\usepackage{color}
\usepackage{mathrsfs}
\usepackage{bm}
\usepackage{amsmath,amssymb,empheq}
\usepackage{mathrsfs}
\usepackage[caption=false]{subfig}
\usepackage{hyperref}
\usepackage[normalem]{ulem}
\usepackage{mathtools}
\usepackage[x11names]{xcolor}
\usepackage[final,nopatch=footnote]{microtype}
\usepackage{orcidlink}

\definecolor{fashionfuchsia}{rgb}{0.96, 0.0, 0.63}
\colorlet{no_so_fashion_purple}{blue!50!red}

\newcommand{\be}{\begin{equation}}
\newcommand{\ee}{\end{equation}}
\newcommand{\ba}{\begin{eqnarray}}
\newcommand{\ea}{\end{eqnarray}}

\newcommand{\nn}{\nonumber}
\newcommand{\half}{\frac{1}{2}}
\newcommand{\fourth}{\frac{1}{4}}

\newcommand{\Lag}{{\cal L}}

\def\half{\frac{1}{2}}

\def\Tr{{\rm Tr}}
\def\rmdiag{{\rm diag}}
\def\etagut{\eta_{\rm GUT}}
\def\epsgut{\epsilon_{\rm GUT}}

\begin{document}
\title{Outcomes of Grand Unified Symmetry Breaking}
\author{Harish Hemming}
\email{hhemming@asu.edu}
\author{Tanmay Vachaspati}
\email{tvachasp@asu.edu}
\author{Anja Wachowitz\,\orcidlink{0009-0005-0920-379X}}
\email{anja.wachowitz@asu.edu}
\affiliation{
Physics Department, Arizona State University, Tempe,  Arizona 85287, USA. 
}

\begin{abstract}
We numerically study the outcome of symmetry breaking motivated by Grand Unified models.
Our results show the formation of biased domain walls for a range of parameters together with magnetic monopoles. The interactions of walls and monopoles leads to novel processes such as the absorption of monopoles on walls, production of monopoles when domain walls annihilate, collapse of walls with magnetic charge that can, with gravity taken into account, produce magnetically charged black holes, and production of a stochastic gravitational wave background from biased domain walls. Our findings open new avenues for probing the grand unification epoch by cosmological observables.
\end{abstract}

\maketitle

\section{Introduction}

The formation of topological defects during phase transitions in the early universe is a generic feature of Grand Unified Theories (GUTs). In particular, magnetic monopoles are of special interest, since the spontaneous symmetry breaking of a single unified gauge group down to the Standard Model gauge group generically leads to their production. 
Because the energy scale of GUTs is far beyond current collider energies, direct experimental tests of GUTs are challenging. Nevertheless, GUTs lead to a number of observable predictions, including proton decay, leptoquarks, and the formation of topological defects.

The production of magnetic monopoles in an early phase transition leads to a disagreement with cosmological observations: Simple estimates predict that monopoles should have been produced in large numbers, yet none have been observed so far. This is known as the monopole problem~\cite{Vilenkin:2000jqa}.

One possible solution of the monopole problem is cosmic inflation. If monopoles are produced before inflation, their number density is exponentially diluted, which explains why they have not been detected.
However, if the relevant phase transition happens after inflation, the number density of monopoles has to be reduced by a different mechanism. One proposed solution involves a second phase transition in which monopoles and antimonopoles get connected by cosmic strings~\cite{Langacker:1980kd,Vilenkin:2000jqa}. The tension of the string pulls them together, leading to a quick annihilation of the monopole and the antimonopole.

Another possibility is that the monopoles interact with domain walls that are produced during the same phase transition~\cite{Dvali:1997sa}. The interaction between monopoles and domain wall has been studied in detail in~\cite{Pogosian:1999zi,Pogosian:2000xv,Vachaspati:2001pw,Pogosian:2001fm,
Pogosian:2002ua,Antunes:2003be,Vachaspati:2003zp,Brush:2015vda,Vachaspati:2006zz,
Dvali:2022rgx,Bachmaier:2023zmq,Senjanovic:2025enc,Bachmaier:2026}. When a monopole comes across a domain wall, it can be trapped on the wall. This allows for a sweeping mechanism: as domain walls move through the universe, they collect monopoles and sweep them away. 

For this mechanism to be viable, the phase transition must produce both monopoles and domain walls. Additionally, the domain walls have to collapse quickly so that the energy density of the universe is not dominated by domain walls. This can be accomplished by introducing a small bias between the vacua. 
Monopoles and antimonopoles are expected to be captured by the domain walls with roughly equal probability, such that the net magnetic charge on the wall mostly cancels. Thus, the collapse of the domain walls can lead to a significant reduction in the monopole abundance.

In this paper we study the interplay between domain walls and monopoles in a thermal phase transition using numerical simulations. Before the phase transition, the fields are in thermal equilibrium. As the system cools below a critical temperature, the potential develops new minima and the symmetry is spontaneously broken. In the resulting phase transition, topological defects form and we study the time evolution of the network.

The minimal GUT is based on an $SU(5)$ symmetry with a scalar field in the adjoint representation, $\Phi$, that gets a vacuum expectation value (VEV). The theory has magnetic monopoles because the final unbroken symmetry contains the $U(1)$ electromagnetic group which has non-trivial first homotopy. Less appreciated is the existence of biased domain walls in the model for a certain range of parameters because $\Phi \to -\Phi$ can be an approximate symmetry of the model~\cite{Pogosian:2000xv,Pogosian:2001fm,Vachaspati:2001pw}. The breaking of this discrete $Z_2$ symmetry implies a cosmological network of domain walls after GUT symmetry breaking. 

The $SU(5)$ GUT model has a very large number of degrees of freedom and so we choose to work with a smaller model that shares the same features of the $SU(5)$ theory. This toy model is based on an $SU(3)$ symmetry and, similarly to $SU(5)$, contains both magnetic monopoles and biased domain walls. We have set up thermal initial conditions for the fields in this model, cooled the system to obtain symmetry breaking, evolved the system and tracked the magnetic monopoles and domain walls.

We find that the evolution of the system depends crucially on the strength of the bias parameters, with more domain walls implying fewer magnetic monopoles, which is consistent with the sweeping scenario of Ref.~\cite{Dvali:1997sa}. Since the walls are biased, they eventually annihilate and lead to the annihilation of magnetic charge as well. A new feature that we observe is the creation of monopole-antimonopole pairs by the collapse of domain walls. In some cases, these pairs are short-lived, but they may also leave behind a residual distribution of monopoles. 
Another possible outcome of grand unification is the creation of magnetically charged black holes.

This work expands upon our recent paper~\cite{Hemming:2026osi} which presented the key results of this project.
In Sec.~\ref{model}
we will describe the 
$SU(3)$
model and the potential we use in detail. In Sec.~\ref{numerical} we explain the numerical setup and in Sec.~\ref{results} we present our results. We discuss the sweeping mechanism in the cosmological context in Sec.~\ref{discussion} and conclude in Sec.~\ref{conclusion}.

\section{The Model}
\label{model}

Our study is motivated by Grand Unified Theories which, besides magnetic monopoles, also produce domain walls during their phase transition in the early universe. The simplest example of such a GUT is $SU(5) \times Z_2$.

However, to simulate an $SU(5)$ gauge theory with an adjoint scalar field we would need to model 120 bosonic fields {($3\times 24$ gauge field components, $24$ scalar components and 24 more variables to enforce the Gauss constraints). We leave out fermions since they are not relevant to the production of topological defects in this scenario. Such a large number of fields is difficult to model in a numerical simulation.

Instead, we consider an $SU(3)$ gauge theory with a scalar field $\Phi$ in the adjoint representation. This model shares several crucial similarities with the above mentioned GUT model: It features similar types of magnetic monopoles and domain walls and it has a similar minimal symmetry breaking pattern to the usual $SU(5)$ to the Standard Model gauge group. The main advantage is that with $SU(3)$ we only have to model $40$ fields ($3 \times 8$ gauge field components, $8$ scalar components and 8 variable to enforce Gauss constraints).

The Lagrangian is given by
\be \label{eq:Lagragian}
\Lag = -\fourth W_{\mu \nu}^a W^{a, \mu \nu} - \half D_\mu \Phi^a D^\mu \Phi^a - V(\Phi),
\ee
where $W_{\mu \nu}^a$ denotes the field strength of the $SU(3)$ gauge field. The adjoint scalar field can be written in components as $\Phi=\Phi^a T^a$, where $T^a$ are the generators of $SU(3)$ in the Gell-Mann convention, normalized as $\Tr (T^a T^b) =\delta^{a b}/2$. The covariant derivative acts on the scalar field as $D_\mu \Phi = \partial_\mu \Phi - i g [W_\mu,\Phi]$, where $g$ is the gauge coupling.

We choose the potential of the scalar field as
\ba
V(\Phi ) &=& - m^2 \Tr (\Phi^2) + \epsilon \,
\sqrt{2} m \, \Tr(\Phi^3) + \lambda \Tr(\Phi^4) \nn \\
&& \hskip 0.5 cm
+  \lambda_6 (\Tr(\Phi^2))^3  + d_6 (\Tr(\Phi^3))^2 
\ea
There are several important things to notice about this potential:
First, for $\epsilon=0$, the potential has an exact $Z_2$ symmetry, which transforms $\Phi \to -\Phi$. This leads to the formation of domain walls when $\Phi$ acquires a vacuum expectation value (VEV). The $\epsilon$-term adds a bias to the domain walls: 
when we turn on a small non-zero $\epsilon$, the domain walls will still percolate during the phase transition and then quickly collapse due to the pressure difference after some duration.

Second, we have included six-dimensional operators. Without them, so including only the mass term and the $\lambda$-term, the potential has an $O(8)$ symmetry. In this case the symmetry breaking pattern is not uniquely determined: Depending on whether the VEV is in $T^3=\rmdiag(1,-1,0)/2$ or $T^8= \rmdiag(1,1,-2)/(2\sqrt{3})$ direction after diagonalizing, the symmetry breaking is given by
\ba
\langle \Phi \rangle \sim T^3:& SU(3) \to U(1) \times U(1),\\
\langle \Phi \rangle \sim T^8:& SU(3) \to U(2).
\ea
We can always diagonalize the VEV to be a linear combination of the diagonal $T^3$ and $T^8$ generators.

The inclusion of six-dimensional terms in the potential lets us choose the symmetry breaking pattern: 
While the term including $\lambda_6$ preserves the $O(8)$ symmetry, the $d_6$-term is responsible for choosing the direction of the VEV.
For $d_6>0$ the VEV is in the $T^3$ direction so that $SU(3)$ is maximally broken to $U(1) \times U(1)$. Instead we choose $d_6<0$, for which the VEV aligns along the $T^8$ direction. This breaks the symmetry $SU(3) \to U(2)$. The requirement that the potential is bounded from below adds two constraints to the parameters, $\lambda_6>0$ and $6 \lambda_6 +d_6 > 0$.

We do not need to include any additional operators in the potential since all operators up to mass dimension six can be expressed in terms of the operators in our potential.
In particular, the following relations hold:
\be
\Tr(\Phi^3) = 3 \det (\Phi),\ \ \Tr(\Phi^4) =\half (\Tr (\Phi^2) )^2
\ee
and
\be
(\Tr(\Phi^2))^3 = 4\Tr(\Phi^6) -\frac{4}{3} (\Tr(\Phi^3))^2.
\ee

The role of the term with $d_6$ can be visualized in the following way: Choosing the adjoint scalar diagonal with components $\Phi=\phi_3 T^3 + \phi_8 T^8$, we can plot the potential in the $\phi_3 - \phi_8$ plane. For $d_6=0$, the $O(8)$ symmetry shows up as the VEV is given by a circle in the $\phi_3 - \phi_8$ plane defined by $\phi_3^2 + \phi_8^2=\eta^2$, where $\eta$ is a constant depending on the parameters in the potential.

The $d_6$-term breaks the $O(8)$ symmetry as can be seen in the component form of the operator
\be
\frac{d_6}{48} \phi_8^2 (-3 \phi_3^2 + \phi_8^2)^2.
\ee
With $d_6 \neq 0$, the vacuum circle turns into 6 discrete minima, as shown on the left side of Fig. \ref{potentialFig}. For $d_6>0$, the operator is minimized by $\phi_8=0$, so that the VEV will be $\sim T_3$. For $d_6<0$, the minima correspond to the three permutations of $\pm \eta T_8$ with
\be 
\eta= 
\sqrt \frac{-2 \lambda + 
 2 \sqrt{ \lambda^2 + 2 (d_6 + 6 \lambda_6) m^2}}{d_6 + 6 \lambda_6}
\ee
Once we introduce a small $\epsilon>0$, $\eta$ gets a small correction and the minima in $+T_8$ permuted directions become energetically lower than the $-T_8$ directions. This is shown on the right side of Fig.~\ref{potentialFig} for $\epsilon=0.15$.

We fix the following parameters for the simulation
\be
\label{eq:constants}
g = 0.5, \, m^2=0.5, \, \lambda = 0.75, \, \lambda_6 = 1.0, \, d_6 = -5.9,
\ee
so that $\eta \approx 1.13$. The potential for these parameters is plotted in Fig.~\ref{potentialFig}.
The parameters were chosen is such a way that $d_6+6\lambda_6 \gtrsim 0$ is small.
With \eqref{eq:constants} and $\epsilon=0$,
the VEV only gets a small correction from the six-dimensional operators, and we can still take $\eta^2 \approx 2 m^2/\lambda$.
When we increase $\epsilon$ in the simulation, $\eta$ gets a small correction proportional to $\epsilon$.
This leads to an energy difference between the $\pm$ vacua of $\Delta V \approx \epsilon \eta^4$.

\begin{figure*}
\includegraphics[width=0.48\textwidth]{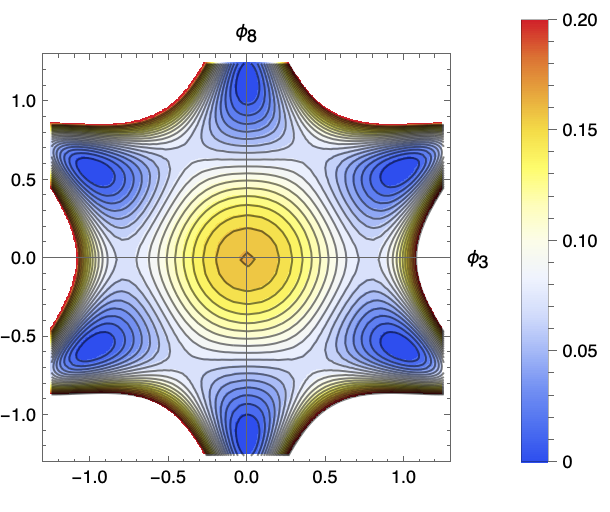} \quad \;
\includegraphics[width=0.48\textwidth]{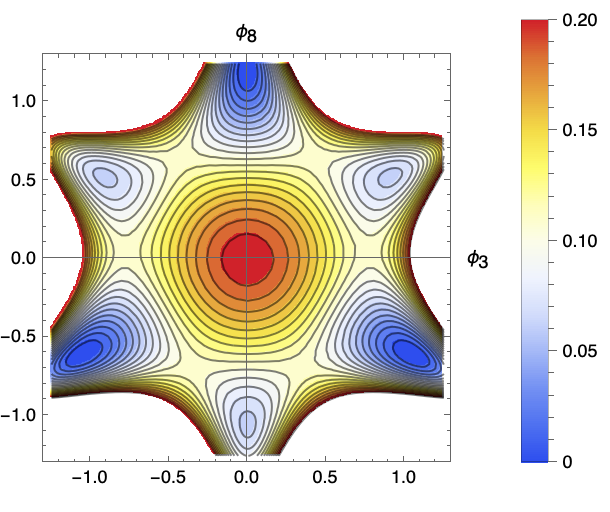}
\caption{Contour plot of potential in $\phi_3-\phi_8$ plane. We added a constant to the potential so that the energy of the vacuum is $0$. {\it Left:} for $\epsilon=0$. {\it Right:} for $\epsilon=0.15$.}
\label{potentialFig}
\end{figure*}

With our choice of parameters and $\epsilon\ne 0$, the symmetry breaking pattern is
\be
SU(3) \to U(2),
\label{eq:symmetrybreaking}
\ee
which leads to the formation of monopoles because of the nontrivial second homotopy group of the vacuum manifold (which is $CP_2$)
\be
\pi_2(SU(3) / U(2))=Z.
\ee

\subsection{The Damping Term}

In the simulation, we evolve the fields according to their equations of motions, which we find from the Lagrangian in Eq. \eqref{eq:Lagragian}.
The equation of motion for the scalar fields $\phi^a$ is
\be
\label{eq:damping_equation}
D_\mu D^\mu \phi^a - \gamma \frac{\phi^b {\dot \phi^b}}{\phi^c \phi^c} \phi^a + \frac{\partial V}{\partial \phi^a} = 0.
\ee
Here, we have added a damping term with damping parameter $\gamma$ which describes the energy loss due to the production of fermions. 

In~\cite{Zhang:2019vsb}, it was explained that a damping term of the form $F_{\rm damp}=-\gamma \phi \partial_t \ln |\phi|$ damps only in radial direction of the field and thus conserves charge, $D_\mu j^\mu=0$. In our case, we take $F_{\rm damp}\sim -\gamma \phi^a \partial_t \ln (\phi^b \phi^b)$. Charge conservation can be shown by considering the $SU(3)$ transformation of the adjoint field given by 
$\Phi \rightarrow g\Phi g^{-1}$ 
where $g \in SU(3)$. The infinitesimal change in each component of the scalar field is given by $\Delta^a \phi_c = -f_{abc} \phi^b$, where the superscript refers to the 8 arbitrary directions by which each field can be perturbed. An expression for the 4-current of the fields can then be computed directly to arrive at the following expression
\be
j^{\mu}_a = \frac{\partial \Lag}{\partial \left(\partial_{\mu}\phi_b\right)}\Delta^{a}\phi_b
          = -f_{abc}\, \phi^b \, D^{\mu}\phi^c .
\ee
Charge conservation can be shown by taking the divergence of the 4-current
\be
D_\mu j^{\mu}_a = -f_{abc} \phi^b \left(D_{\mu}D^{\mu}\phi^c \right)= 0
\ee
where Eq.(\ref{eq:damping_equation}) was used to substitute for the second covariant derivative. Furthermore, the form of the damping term also preserves gauge invariance since $\phi^b \dot{\phi^b}$ 
can be expressed as $\partial_0(\phi^b \phi^b)/2$ and $\phi^b \phi^b$ is proportional to $\Tr(\Phi^2)$
which is gauge invariant.

The equations of motion for the gauge fields are
\be
[D_\mu , W^{\mu\nu} ] = j^\mu = ig [\Phi, D^\nu\Phi].
\label{eq:Wiaeq}
\ee

\subsection{The Monopole}
\label{themonopole}

Since $\pi_2(SU(3)/U(2))=Z$, the symmetry breaking gives rise to monopoles, discussed in~\cite{Bais:1978yh,Wilkinson:1978zh}.
The ansatz for the $SU(3)$ monopole is the field configuration
\be
\Phi_{\rm m} ({\bf x}) = \eta \left ( f_m (r)  \frac{\sqrt{3}}{2} {\hat r} \cdot {\hat \tau} - g_m(r) \frac{1}{2} T_8 \right )
\label{eq:Phim}
\ee
where the angular dependence is in the unit radial vector $\hat r$ and ${\hat \tau}=(T_1,T_2,T_3)$ are three generators of $SU(3)$ that define an $SU(2)$ subgroup. The profile functions $f_m$ and $g_m$ have to fulfill the following boundary conditions: For the solution to be regular, we require $f_m(0)=0$. Notice that for $g_m(0)$ there is no such restriction because all the angular dependence is in the $f_m$ term. Instead, we have $g_m'(0)=0$. At large distances, $f_m(\infty)=g_m(\infty)=1$ so that $\Phi(\infty)=\eta \, \rmdiag(1,-2,1)/2\sqrt{3}$ is in one of the minima. 

This monopole ansatz can be rotated in $SU(3)$ space. In general, any $SU(3)$ monopole can be described by a winding in an $SU(2)$ subgroup corresponding to the profile $f_m$ and an additional profile $g_m$ that corresponds to the generator that commutes with the $SU(2)$ winding subgroup.

The solution for the profile functions 
in the global $SU(3)$ case is found numerically and is shown in Fig.~\ref{fig:mprofiles} for $\epsilon=0$. For nonzero but small $\epsilon \ll 1$, the profile functions look almost the same. We expect the local monopole profile to be similar. Fig.~\ref{fig:monopoletrphisq} shows the radial dependence of ${\rm Tr} (\Phi^2 )= \eta^2(3 f_m^2(r) +g_m^2(r))/8$ calculated from the monopole profile functions. While in the vacuum, ${\rm Tr} (\Phi^2) \approx 0.64$, this plot shows that in the monopole center ($r=0$) ${\rm Tr} (\Phi^2) \approx 0.16$ is much smaller. Thus, to find monopoles in the simulation, the first filtering will look for points where ${\rm Tr} (\Phi^2)$ is small.

\begin{figure}
\includegraphics[width=0.45\textwidth]{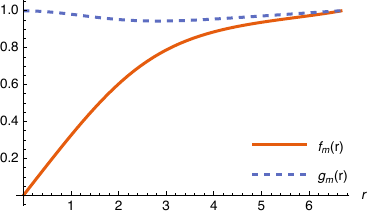}
\caption{Monopole profile functions $f_m(x)$ and $g_m(x)$, $\epsilon=0$.}
\label{fig:mprofiles}
\end{figure}

\begin{figure}
\includegraphics[width=0.45\textwidth]{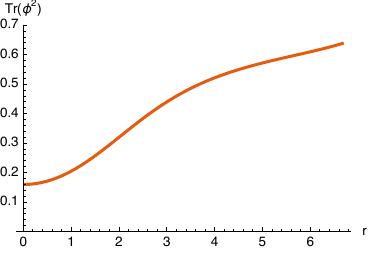}
\caption{${\rm Tr} (\Phi^2)$ for monopole profile, $\epsilon=0$.}
\label{fig:monopoletrphisq}
\end{figure}

\begin{figure}
\includegraphics[width=0.45\textwidth]{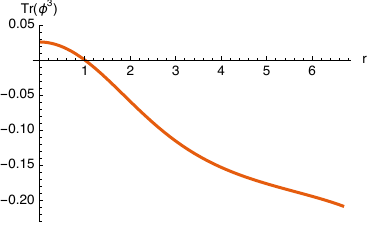}
\caption{${\rm Tr} (\Phi^3)$ for monopole profile, $\epsilon=0$.}
\label{fig:monopoletrphicube}
\end{figure}

To estimate the number density of monopoles that we expect to produce in the numerical simulation, we need to know the correlation length at the phase transition $\xi$. Let us denote the probability to generate a $\pm1$ winding in a cube whose size is $\xi$ by $p_m$.
Then, the number density of monopoles plus antimonopoles in the simulation is $n_m = p_m /\xi^3$.

To estimate $p_m$, we can repeatedly take 8 random points from the vacuum manifold, arrange them on the vertices of the cube of size $\xi$, and calculate their winding. To pick random points on the vacuum manifold, we follow the derivation from~\cite{Ng:2008mp}. The vacuum manifold for the symmetry breaking given in Eq. \eqref{eq:symmetrybreaking} is
\be
SU(3)/U(2) \cong \mathbb{C}P^2.
\ee
The space $\mathbb{C}P^2$ consists of complex vectors $Z=(z_1,z_2,z_3)^T$ where rescaled vectors are identified with each other,
\be
(z_1,z_2,z_3) \cong \kappa (z_1,z_2,z_3), \; \kappa \in \mathbb{C}\backslash \{0\}.
\ee
We can parametrize $Z \in \mathbb{C}P^2$ as
\be
Z^T = ( \sin \bar{\theta} \cos \bar{\phi} \; e^{i \alpha}, \sin \bar{\theta} \sin \bar{\phi} \; e^{i \beta},  \cos \bar{\theta} ),
\ee
where $0 \leq \bar{\theta}, \bar{\phi} \leq \pi/2$ and $0 \leq \alpha, \beta \leq 2 \pi$.
Because of the $SU(3)$-invariant measure on $\mathbb{C}P^2$, we draw from a uniform distribution $0 \leq \sin^4 \bar{\theta} \leq 1$, $0 \leq \sin^2 \bar{\phi} \leq 1$, $0 \leq \alpha \leq 2 \pi$ and $0 \leq \beta \leq 2 \pi$.
The magnetic charge inside a lattice constructed from 8 randomly chosen $Z$-vectors is
\be
Q = \frac{1}{2\pi} \sum_{ijkl} \arg \left( (Z_i^\dagger Z_j) (Z_j^\dagger Z_k) (Z_k^\dagger Z_l) (Z_l^\dagger Z_i) \right),
\ee
where the sum is over the six faces of the lattice cube.

From drawing 8 random $Z$'s $10^6$ times and calculating the winding in their cell, we find approximately $65.2 \%$ cells with no magnetic charge, $17.1\%$ with $+1$ charge, $17.1\%$ with $-1$ charge, $0.2\%$ with $+2$ and $0.2\%$ with $-2$.
Thus, we estimate the probability to get a $+1$ winding is $\approx 17\%$ (also $17\%$ for $-1$ winding).
The probability for a $\pm 1$ winding is $p_m \approx 0.34$.

The correlation length $\xi$ depends on details of the phase transition and has been the subject of discussion~\cite{Kibble:1976sj,Zurek:1996sj,Vachaspati:2006zz} 
For a second order transition an order of magnitude estimate is $\xi \simeq (g^2 T)^{-1}$ where $T\simeq \eta$ is the temperature at the phase transition.

\subsection{The Domain Wall}

The breaking of the exact ($\epsilon=0$) or approximate ($\epsilon \ll 1$) $Z_2$-symmetry produces domain walls. In case of $\epsilon=0$, the domain walls are topological and stable, while for $\epsilon \neq0$ the walls are non-topological and unstable. Similar domain wall solutions have been discussed previously in~\cite{Pogosian:2001fm,Pogosian:2000xv,Pogosian:2002ua,Pogosian:2001pq,Antunes:2003be,Vachaspati:2003zp}.

For $\epsilon=0$, the potential has 6 minima corresponding to the three permutations of $\pm T^8$. The domain walls interpolate between minima of opposite sign. A small $\epsilon$ introduces a bias in the energy of those minima.
The domain walls with lowest energy interpolate between two minima that are next to each other in the $\phi_3$-$\phi_8$ potential. One minimum will be in $+T^8$ direction and the other in $-T^8$ direction with permutation.

For example, there is a domain wall solution that interpolates between $\Phi(x=-\infty) = \eta \, \rmdiag(1,-2,1)/2\sqrt{3}$ and $\Phi(x=\infty) = -\eta \, \rmdiag(-2,1,1)/2\sqrt{3}$. The ansatz for the domain wall configuration is
\be
\Phi_{\rm dw} (x) = g_w(x) T_3 + f_w(x) T_8,
\ee
with boundary conditions $f_w(\pm \infty) = \pm \eta/2$, $f_w(0)=0$, $g_w'(0)=0$ and $g_w(\pm \infty)=\eta \sqrt{3}/2$.
The solution for $f_w(x)$ and $g_w(x)$ is solved numerically and shown in Fig.~\ref{dwprofiles}.

Notice that at the center of the domain wall, the field is in the $T^3$ direction. Thus, the domain wall is a layer where the symmetry is broken to $U(1) \times U(1)$.

What is $\Tr (\Phi^2)$ at the center ($x=0$) of the domain wall? From the profile functions obtained numerically, we calculate $\Tr (\Phi^2)=0.34$ at the center. Thus, when we look for monopoles at points where $\Tr (\Phi^2)$ is small, we will also encounter domain walls and have to filter them out.

To estimate how many walls are produced during the phase transition, we subdivide the lattice into smaller cubes with size $\xi$, where $\xi$ is the correlation length of the field at the phase transition. Now we can consider each of the cubes as made up of the 8 uncorrelated $\Phi$ values at the vertices.
To have no domain wall passing through one of the cubes, $\Tr(\Phi^3)$ at each of the 8 points of the cube should have the same sign. This has 
probability $2 \cdot (1/2)^8=1/128$ since for one point it is $1/2$ to have the sign $\pm$ for $\epsilon=0$.
Thus, the probability to have a domain wall pass through the cube is $1-1/128=0.992$. 

So just after the phase transition, if $\epsilon$ is very small, in almost all cells with volume $\xi^3$ there will be a domain wall, but only some of those cells will also have a monopole winding (see Sec.~\ref{themonopole}). That means windings are always created near domain walls, or even on domain walls. The latter case corresponds to magnetically charged domain walls, not from the sweeping 
mechanism, but created during the phase transition. When there are several opposite windings on the same domain wall, they can be canceled or at least reduced very quickly. Thus, we expect the most stable domain wall network in the case with $\epsilon \to 0$ to be best at canceling the magnetic charges.

\begin{figure}
\includegraphics[width=0.45\textwidth,angle=0]{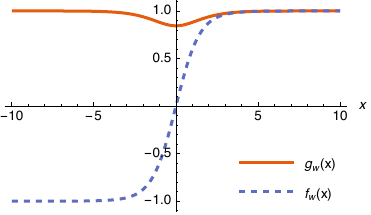
}
 \caption{Domain wall profile functions $f_w(x)$ and $g_w(x)$.}
\label{dwprofiles}
\end{figure}

\section{Numerical Setup}
\label{numerical}

Our simulation is performed on a $300^3$ lattice grid with periodic boundary conditions using parameters $dx=0.5$ and $dt={dx}/{3}$ along with parameters for the potential} given in Eq~\eqref{eq:constants}. We work in the temporal gauge $\left(W^a_0 = 0\right)$ giving a total of 24 gauge fields and 8 scalar fields that are each evolved at every time step using a position Verlet leapfrog algorithm, where second-order central differences were used for computing derivatives. The corresponding equations of motion for the scalar and gauge fields are given by 
\be
\begin{split}
\label{eq:scalarEq}
\partial_0\partial^0\phi^a &= \nabla^2 \phi^a + gf_{abc}W^b_i\partial_i\phi^c + \frac{1}{2}gf_{abc}\phi^c\Gamma^b \\
&\quad + \frac{1}{4}g^2f_{ace}f_{dbe}\phi^bW^d_iW^c_i + \frac{1}{2}m^2\phi^a \\
&\quad - \frac{3}{8}\lambda\phi^a\phi^b\phi^b - \frac{3}{4}\epsilon d_{abc}\phi^b\phi^c \\
&\quad - \frac{3}{4}\lambda_6 \phi^a\left(\phi^b\phi^b\right)^2 \\
&\quad - \frac{3}{8}d_6 \left(d_{abc}\phi^b\phi^c \right)\left(d_{def}\phi^d\phi^e\phi^f \right)
\end{split}
\ee

\be
\begin{split}
\label{eq:gaugeEq}
\partial_0\partial^0 W^a_i &= \nabla^2 W^a_i - \partial_i \Gamma^a -2gf_{abc}W^c_j\partial_jW^b_i \\
&\quad - gf_{abc}W^b_i\Gamma^c + gf_{abc}W^c_j\partial_iW^b_j \\
&\quad -\frac{1}{2}gf_{abc}\phi^b\partial_i\phi^c - \frac{1}{4}g^2f_{abe}f_{cde}\phi^b\phi^dW^c_i \\
&\quad -g^2f_{aed}f_{ebc}W^b_jW^d_jW^c_i 
\end{split}
\ee
where $d_{abc}$ are the symmetric structure constants that arise from non-trivial anti-commutation relations in $SU(3)$. We have also introduced 8 additional fields that we define as $\Gamma^a = \partial_iW^a_i$ to enforce Gauss constraints as described in Ref.~\cite{Vachaspati:2016abz}. The corresponding equations of motion for $\Gamma^a$ are given by
\ba
\partial_0\Gamma^a &=& \partial_i\partial_0W^a_i - g_{p2} \bigl (\partial_i\partial_0W^a_i - gf_{abc}W^c_i\partial_0W^b_i \nn \\
&& \hskip 3.5 cm
+ g f_{abc}\phi^b\partial_0\phi^c \bigr )
\label{eq:constraintEq}
\ea
where $g_{p2}$ is a numerical parameter that helps with numerical stability and we choose it to be $g_{p2}=0.81$.

\subsection{Initial Conditions}

Thermal initial conditions for the simulation are implemented by randomly sampling
the Fourier modes for the fields ($\phi^a(\vec{k}), W^a_i(\vec{k})$) from a Gaussian distribution with variance given by the Bose-Einstein (BE) distribution,
\be
\sigma_k^2 = \frac{2}{\omega_k} \frac{1}{e^{\omega_k/T} - 1}
\label{variance}
\ee
The BE 
distribution involves two parameters, one is the temperature, $T$, and the other is
the mass parameter in the dispersion relation $\omega_k = \sqrt{k^2+\mu^2}$. For the 
purpose of the initial conditions we choose $\mu=1=T$  and set $\mu=0$, $T=1$ for the gauge fields. Once the initial field configuration in Fourier space is sampled, we perform a Fast Fourier Transform to obtain the corresponding field configuration in position space. The Gauss constraints are initialized by simply computing $\Gamma^a = \partial_iW^a_i$. A similar prescription to initialize the velocities of the field is not as straightforward since the Gauss constraints given by the expression in parentheses of Eq. (\ref{eq:constraintEq}) is required to vanish at all times and an arbitrary field configuration generated from a random distribution is not expected to satisfy such a constraint. We avoid this issue by setting all time derivatives of the field to $0$. The initial conditions are therefore ``half'' thermal in nature and we include a factor of 2 in the variance in \eqref{variance} to compensate for this. Once we have the fields, we evolve them according to the
equations of motion in Eq.~(\ref{eq:scalarEq}), Eq.~(\ref{eq:gaugeEq}) and Eq.~(\ref{eq:constraintEq}). The interactions among the
fields result in their thermalization (at a temperature different from $T=1$) after a short time 
which we take to be when the kinetic and gradient energies have stabilized to a constant value. Note 
that the field interactions effectively give a thermal mass to the $\Phi$ field and
there is no need to include a thermal term in the potential during evolution. To then obtain symmetry
breaking, we cool the system by turning on a damping term with damping coefficient
$\gamma =0.6$ in Eq.~(\ref{eq:damping_equation}).

\subsection{Algorithm for Identifying Monopoles}
\label{monoalgo}

The first step in our algorithm involves mapping out regions in the lattice where ${\rm Tr}(\Phi^2)$ is below a certain threshold as this ensures we are close to the vicinity of monopoles, avoiding the computationally expensive task of checking for monopoles in every lattice cell. Motivated by the profile function of the monopole in Fig.~\ref{fig:monopoletrphisq}, we restrict ourselves to locations where $\Tr(\Phi^2) < 0.3$ as this indicates regions where the potential energy is high and likely corresponds to locations of defects.
 
\begin{figure}[ht]
\includegraphics[width=\linewidth, keepaspectratio]{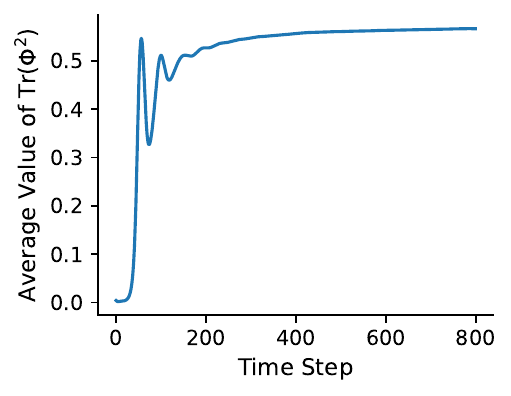}
\caption{Plot of averaged value of $\Tr(\Phi^2)$ as a function of time.
}
\label{fig:Phi2_vs_Time}
\end{figure}

Furthermore, we only search for monopoles once the average value of $\Tr(\Phi^2)$ over the lattice reaches 0.5, indicating that most of the field has settled close to its true vacuum where $\Tr(\Phi^2) = 0.64$. Fig.~\ref{fig:Phi2_vs_Time} shows that this occurs around $t=150$ and this is when we start searching for monopoles using our algorithm.
 
Once we have filtered for points satisfying $\Tr(\Phi^2) < 0.3$, we further isolate points where $\Tr(\Phi^2)$ is a local minimum when compared with neighboring points centered around a $(3^3)$ sub-lattice. The local minima are most likely to be close to the monopole center.

Once the local minima of $\Tr(\Phi^2)$ have been identified, we check if any of them correspond to domain walls or monopoles that are about to be swept up by domain walls. 
We rule out such situations by constructing $(7^3)$ sub-lattices around those points and checking that the sign of $\Tr(\Phi^3)$ is the same at all 8 vertices of each sub-lattice. The size of our sub-lattice was chosen to capture the inner core of the monopole. (We define the monopole core as the region where $\Tr(\Phi^3)$ has the opposite sign with respect to its sign in the asymptotic region. See Fig.~\ref{fig:monopoletrphicube}.) Hence any sign flip of $\Tr(\Phi^3)$ along any edge of the sub-lattice will only arise from a domain wall passing through the sub-lattice. 
Furthermore, once a sub-lattice has been identified using this procedure, the sign of $\Tr(\Phi^3)$ on the boundary reflects 
the asymptotic vacuum ($+ T^8$ or $-T^8$).

The next step involves identifying candidate cells inside the sub-lattice to compute the winding of the field and detect the presence of monopoles. We avoid calculating the winding throughout the sub-lattice since we have found that the accuracy of our algorithm is highly sensitive to how close we are to the monopole center. We restrict ourselves to cells where $f_m$ in Eq. (\ref{eq:Phim}) is below some small threshold ($f_m^2 < 0.08$) as this indicates that we are close to the center of a monopole. To compute the value of $f_m$ at each point, we solve the simultaneous equation derived from finding the expression for ${\rm Tr}(\Phi_{\rm m}^3)$ and ${\rm Tr}(\Phi_{\rm m}^2)$ using the monopole ansatz in Eq.~(\ref{eq:Phim}) given by,
\be
\Tr (\Phi_m^2) = \frac{\eta^2}{8}(3 f_m^2+g_m^2 ), \ \ 
\Tr (\Phi_m^3) = \frac{\eta^3}{32\sqrt{3}} g_m (9 f_m^2 - g_m^2). \nn
\ee
A unique solution for $g_m$ and $f_m$ is obtained by imposing the constraint that the sign of $g_m$ is the same as the sign of the vacuum outside the monopole core (sign$(g_m)=\pm$ for ${\rm Tr}(\Phi^3) = \mp$ in the vacuum) and $f_m^2 < g_m^2$. 

The points where $f_m$ is small are close to the center of the monopole.
Denote one such point 
by ${\bf x}_*$. Let $U_*$ diagonalize ${\tilde \Phi} ({\bf x}_*)$, where the tilde denotes $\Phi$ multiplied by the sign of $g_m$ at that point. The diagonal matrix, $D_* \equiv U_* {\tilde \Phi}_* U_*^\dag$ has two eigenvalues that are smaller than the third as it is approximately
proportional to $T_8$ or to $T_8$ but with permuted diagonal entries. These smallest eigenvalues identify the unbroken SU(2) block. Note that multiplying $\Phi$ by the sign of $g_m$ before diagonalizing ensures that this remains true even in the case where the vacuum outside the monopole core is in the $-T_8$ sector instead of $+T_8$ sector.

We also observe that this method correctly identifies the SU(2) block that encodes the winding of the field even when the closest point to the monopole center is outside the monopole core, such as may occur at early times.

We denote the generators of the SU(2) group that commute with $D_*$ by
${\vec \Sigma}$. Then the generators of the unbroken SU(2) group are 
${\vec \tau} \equiv U_*^\dag {\vec \Sigma} U_*$. Once we have ${\vec \tau}$, we
construct the radial unit vector in 
Eq.~(\ref{eq:Phim}) in the 8 adjacent sub-lattices that share the vertex at ${\bf x}_*$,
\be
{\hat r}_{i,j,k} = \Tr ( {\vec \tau} \Phi_{i,j,k})/ | \Tr ( {\vec \tau} \Phi_{i,j,k})|.
\ee
In a given cell, we now have 8 ${\hat r}$ vectors, one at each vertex of the cell.
These 8 ${\hat r}$ vectors define a mapping from the surface of the cell to a two
sphere. The next step is to determine if the mapping is topologically 
non-trivial~\cite{Leese:1990cj}.

Consider a ``triangular plaquette'' of the cell, for example formed by the vertices
$(i,j,k)$, $(i+1,j,k)$ and $(i,j+1,k)$. The ${\hat r}$ vectors at these points maps
the triangular plaquette to a spherical triangle on the two sphere. The area of this 
spherical triangle is given by
\be
{\cal A} = 2\, \tan^{-1} \left ( \frac{{\hat r}_1 \cdot {\hat r}_2 \times {\hat r}_3}{1+r_{12}+r_{23}+r_{31}}
 \right )
\ee
where $r_{ab} = {\hat r}_a \cdot {\hat r}_b$ and $a,b =1,2,3$ refer
to the three vertices of the spherical triangle and the triangular plaquette $abc$ is 
oriented with areal vector pointing out of the cell.
Three points on a two sphere define two spherical triangles and we
always take $A$ to be the area of the smaller triangle: $-2\pi \le A \le 2\pi$. 
Next we sum the areas of the 12 spherical triangles obtained
as maps from the 12 triangular plaquettes (``tp'') of the cell to get the topological
winding
\be
w = \frac{1}{4\pi} \sum_{\rm tp} A_{\rm tp}.
\ee
The magnitude of $w$ gives the charge while the sign of $w$ identifies if the cell contains a monopole or anti-monopole. The total number of monopoles at each time step $t$ is then computed using $N_m(t) = \Sigma \lvert w\rvert$.

\subsection{Domain Wall Area}
The algorithm for counting the total domain wall areal density follows naturally from the monopole algorithm described in the previous section. Once monopole containing $(7^3)$ sub-lattices are identified, the remaining lattice cells outside those sub-lattices become candidate cells in which domain walls may be present. In order to detect the presence of domain walls in each cell, we simply count the number of sign flips of ${\rm Tr}(\Phi^3)$ along all 12 edges in each cell and identify a 
domain wall in that cell if there is at least one sign flip along any of the 12 edges. The total number of such cells in the lattice is then summed and multiplied by the area of a plaquette ($dx^2$) to get the total area
of the domain walls at each time step. Division by the volume of the lattice gives us the areal density as a function of time.

\section{Results}
\label{results}

\begin{figure*}
\includegraphics[width=0.48\textwidth,trim={1.8cm 1.2cm 3.4cm 1.5cm},clip]{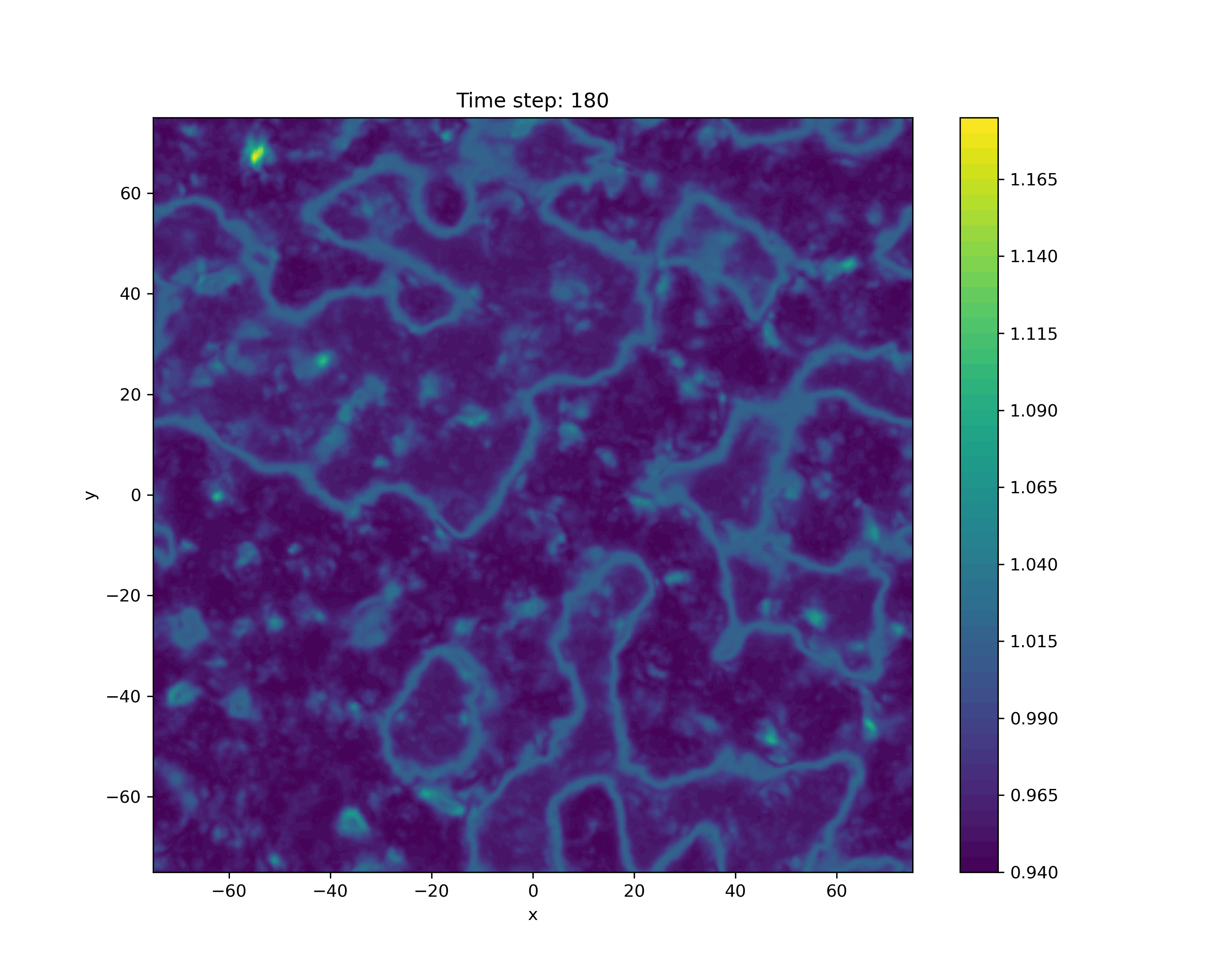} \quad \;
\includegraphics[width=0.48\textwidth,trim={1.8cm 1.2cm 3.4cm 1.5cm},clip]{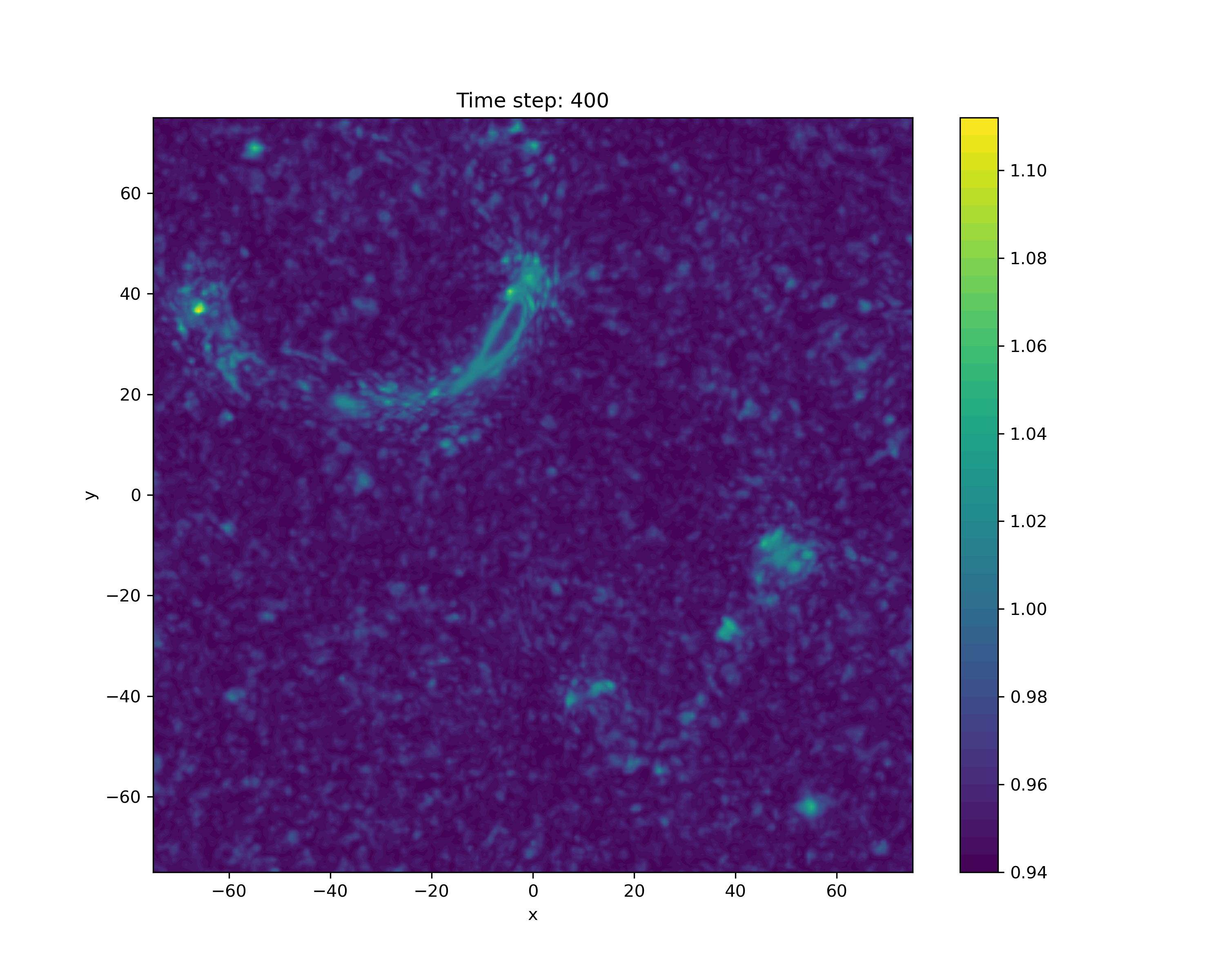}
\includegraphics[width=0.48\textwidth,trim={1.8cm 1.2cm 3.4cm 1.5cm},clip]{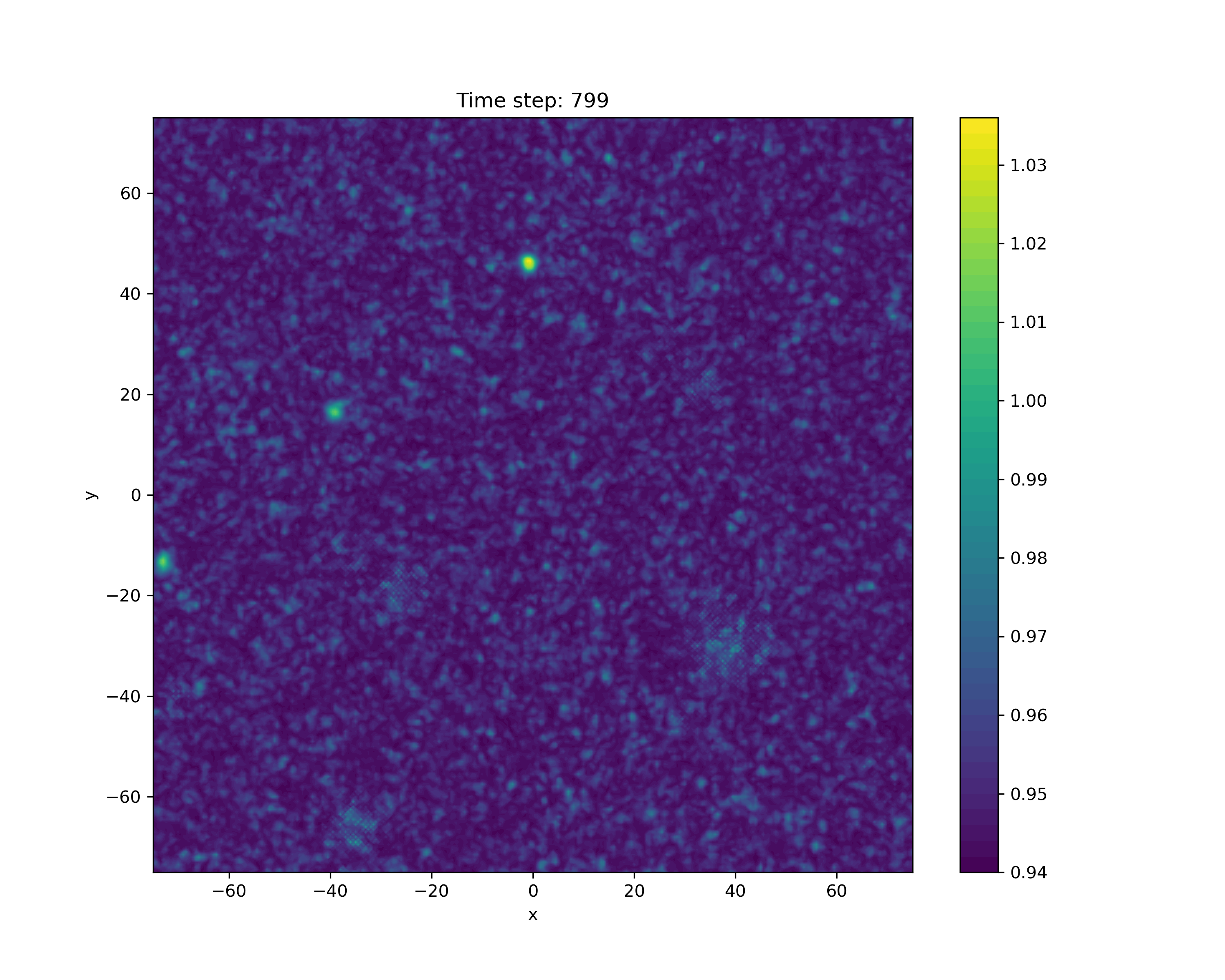}
\caption{Snapshots of the potential energy density for the slice at $z=0$ at three intermediate time steps ($t=180,400,799$). This is for the simulation with $\epsilon = 0.02$.}
\label{fig:potentialenergy}
\end{figure*}

In our simulations, we see both the formation of domain walls and monopoles. In Fig.~\ref{fig:potentialenergy}, we plot snapshots of the potential energy density for a run with $\epsilon=0.02$ at one 2D-slice ($z=0$) and three different time steps $t=180,400,799$.   
At time step $t=180$, we can see the domain wall network that forms right after the phase transition. Some of the domain walls, that have a very small circular shape, are about to collapse. The point like increases in the potential energy are either magnetic monopoles or just fluctuations from the phase transition. When the larger domain walls collapse, they sweep all monopoles that are inside their volume.

At time step $t=400$, almost all domain walls have collapsed. In the slice $z=0$ there is only one domain wall left which is very small and about to collapse. At the rest of space, the field is already in its true vacuum, $\sim +T_8$. There are also a few spots with larger potential energy, which are possible magnetic monopoles. They could have been formed either at the phase transition, or more likely, from the collapse of magnetically charged domain walls. After the entire domain wall network has collapsed, there are only some magnetic monopoles left. Even at time step $t=799$, we can clearly see several magnetic monopoles on this $z=0$ slice.

\begin{figure*}
\centering
\subfloat[$t=200$]{%
  \includegraphics[width=0.32\textwidth]{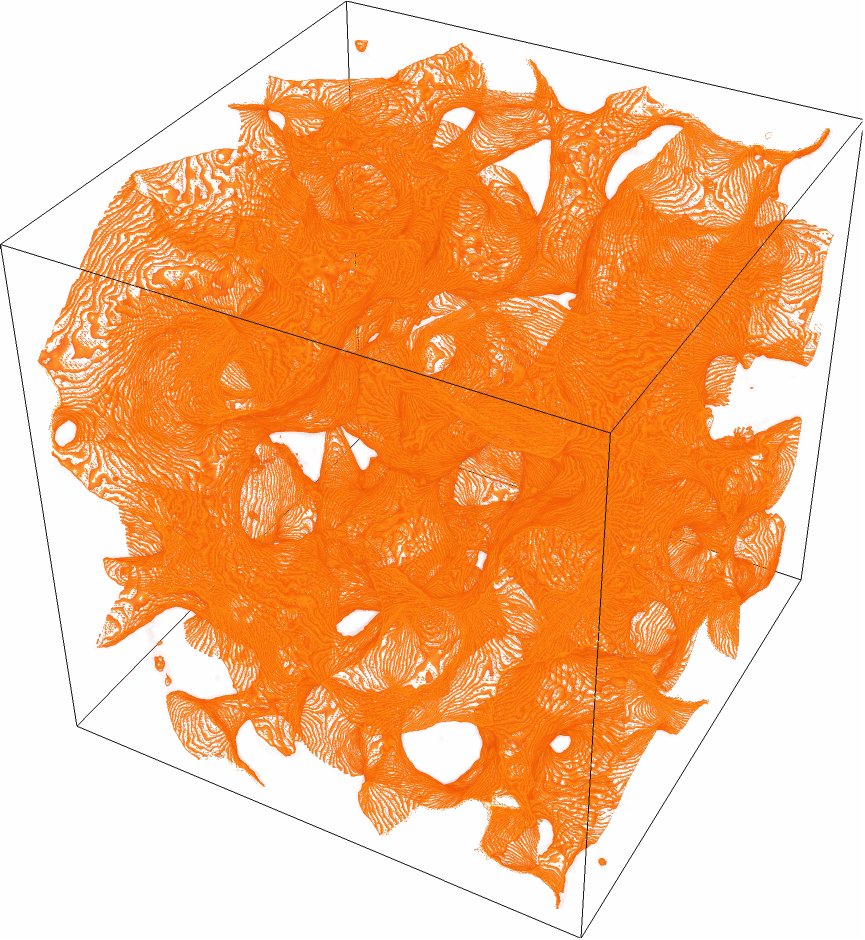}}
\hfill
\subfloat[$t=300$]{%
  \includegraphics[width=0.32\textwidth]{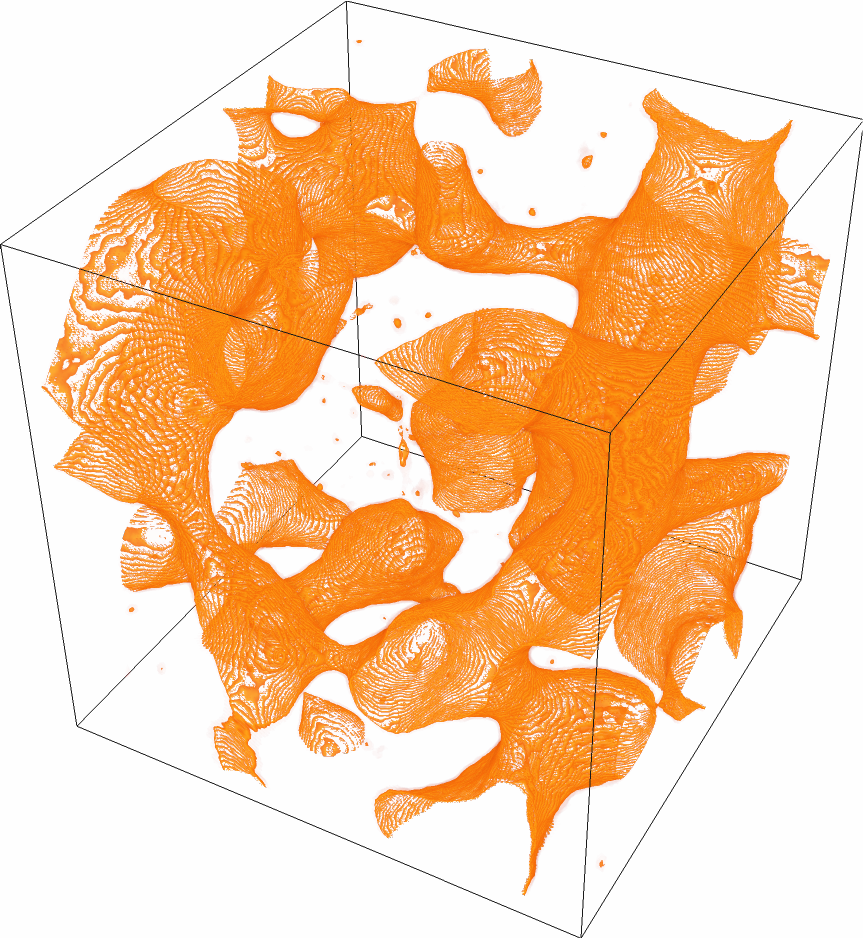}}
\hfill
\subfloat[$t=400$]{%
  \includegraphics[width=0.32\textwidth]{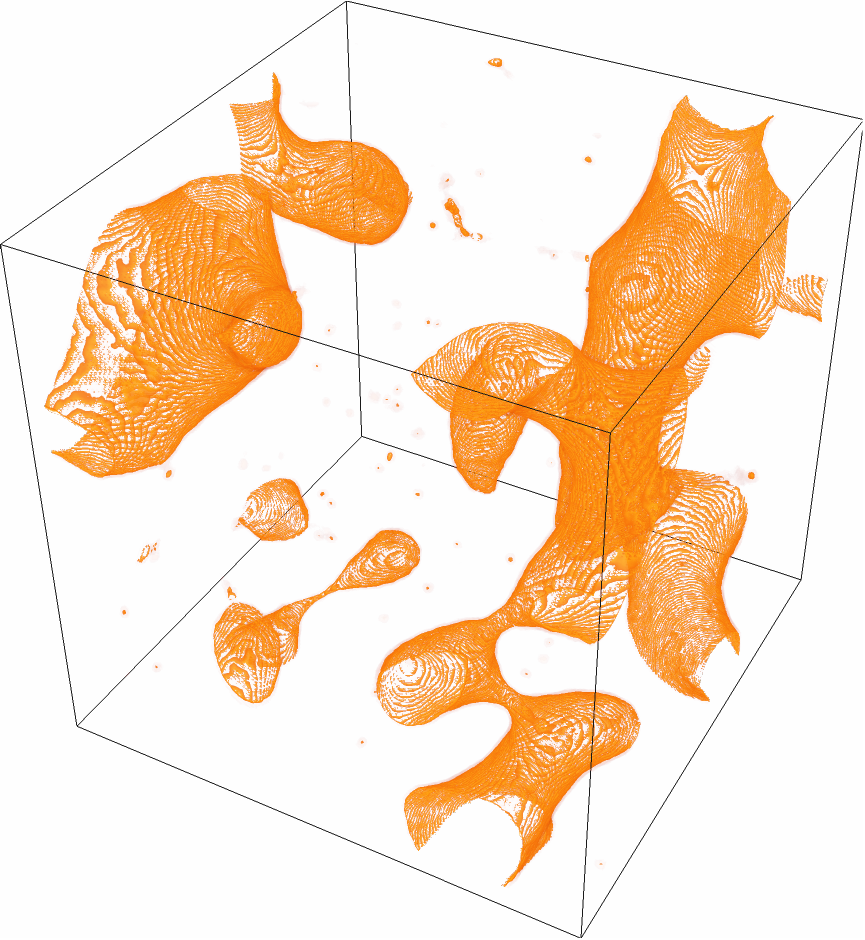}}

\vspace{1ex}

\subfloat[$t=500$]{%
  \includegraphics[width=0.32\textwidth]{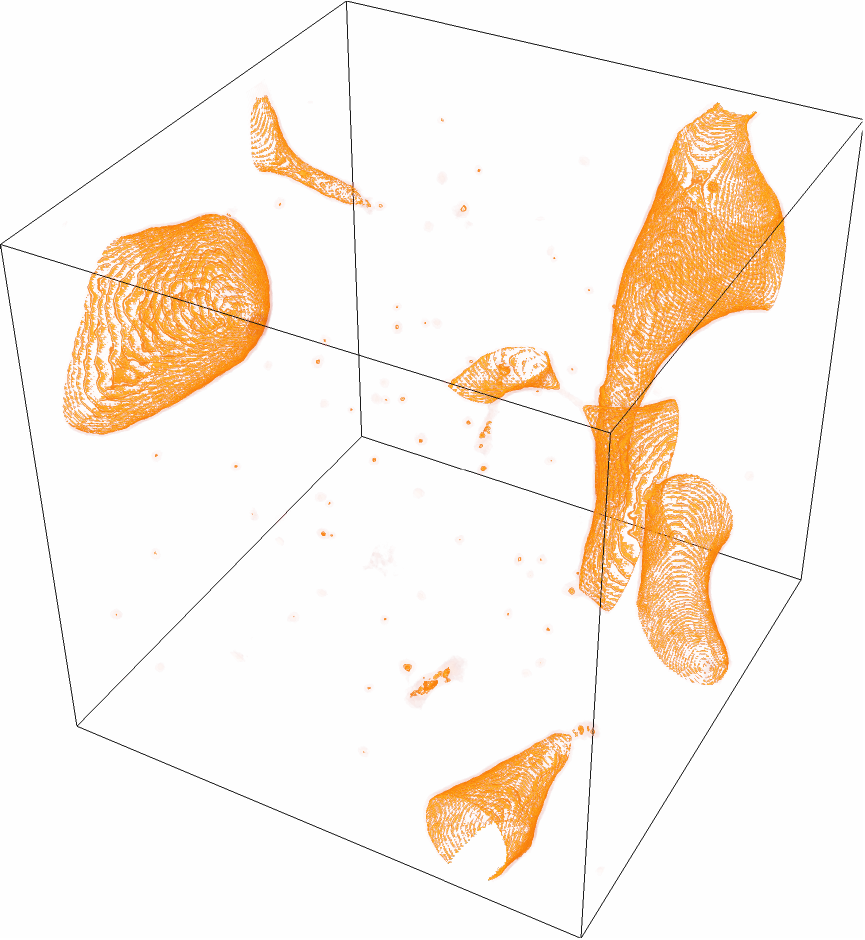}}
\hfill
\subfloat[$t=600$]{%
  \includegraphics[width=0.32\textwidth]{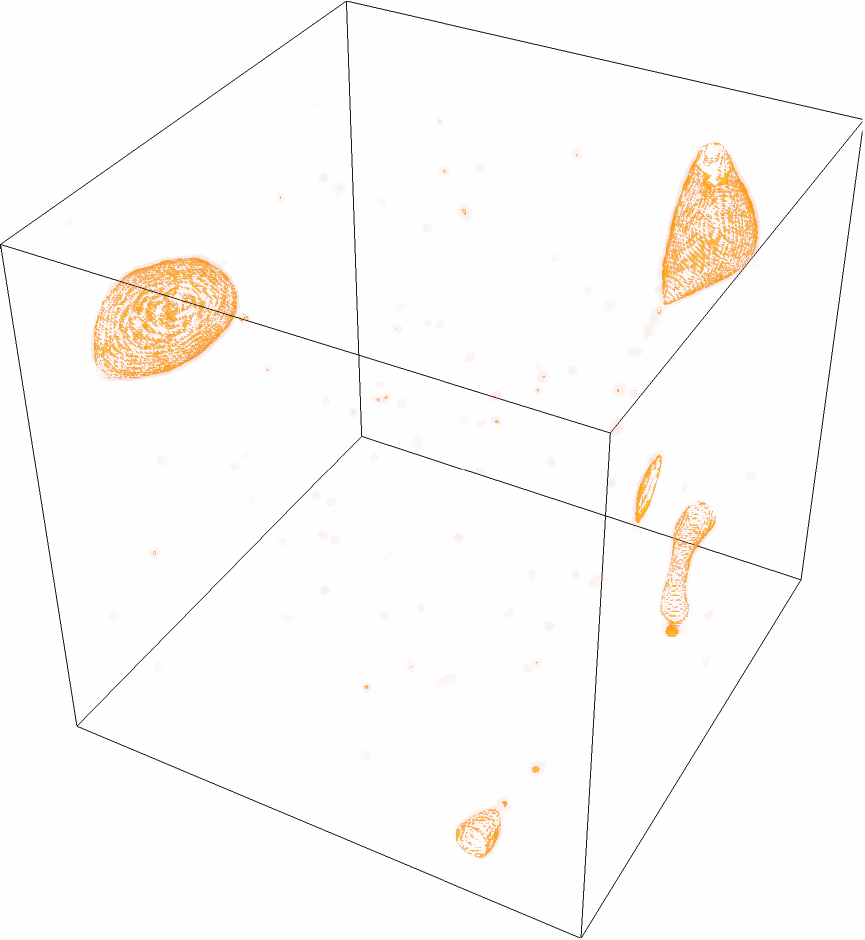}}
\hfill
\subfloat[$t=700$]{%
  \includegraphics[width=0.32\textwidth]{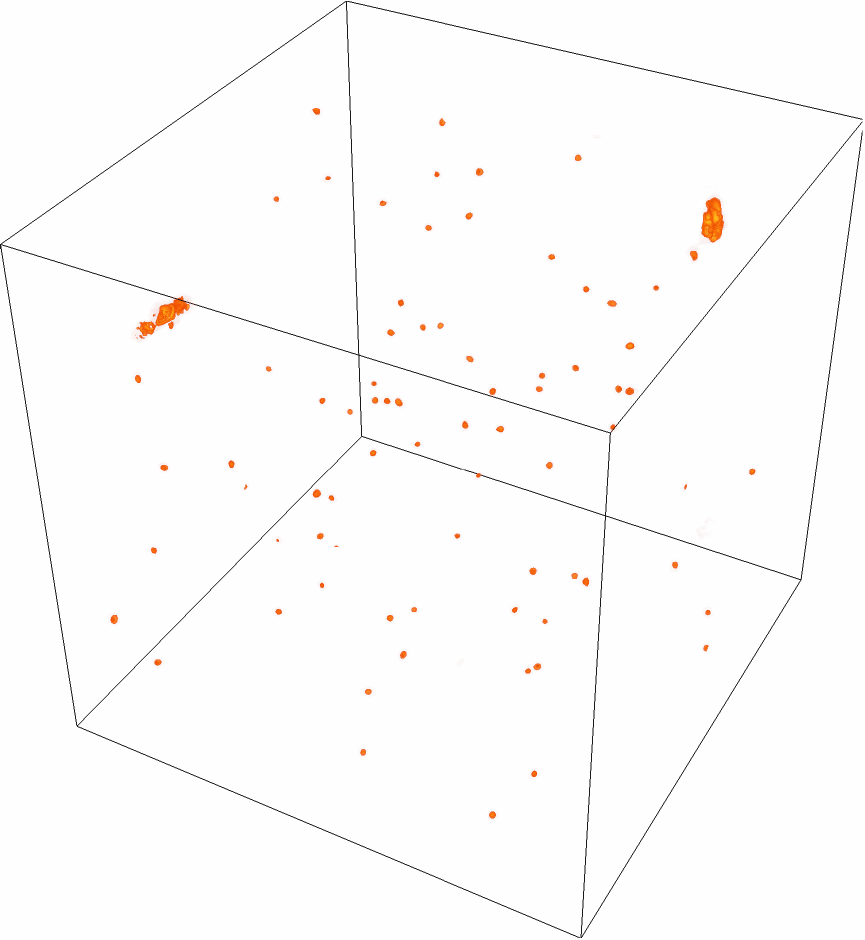}}

\vspace{1ex}

\hspace*{\fill}
\subfloat[$t=800$]{%
  \includegraphics[width=0.32\textwidth]{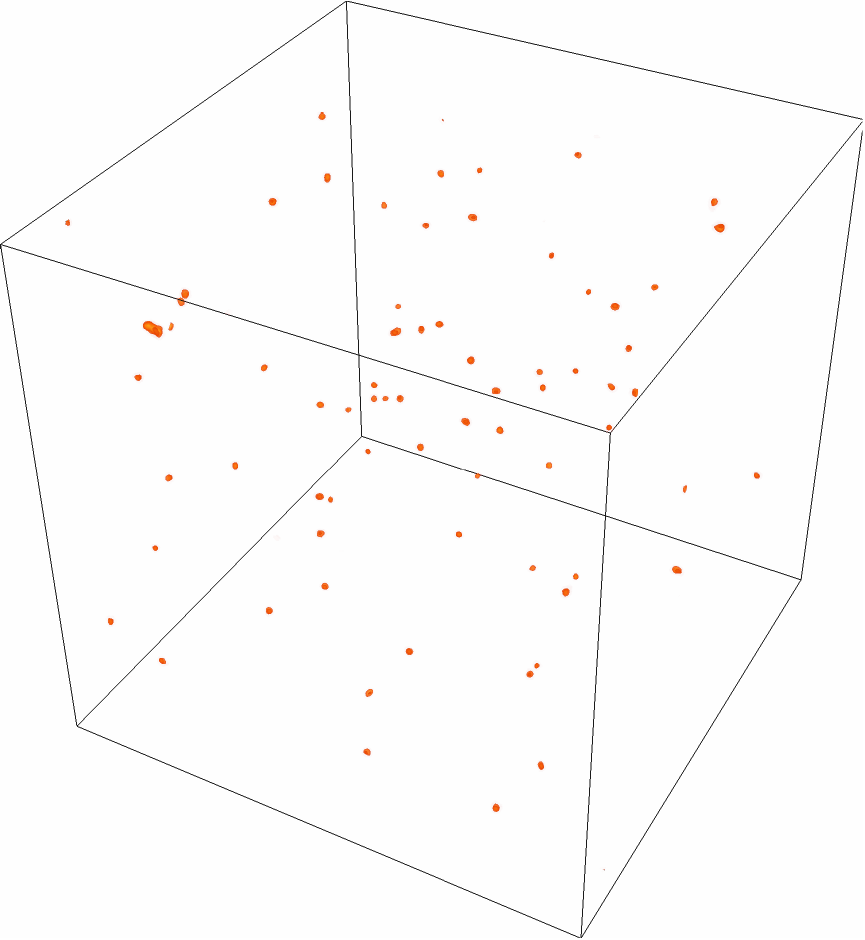}}
\hspace{0.02\textwidth}
\subfloat[$t=900$]{%
  \includegraphics[width=0.32\textwidth]{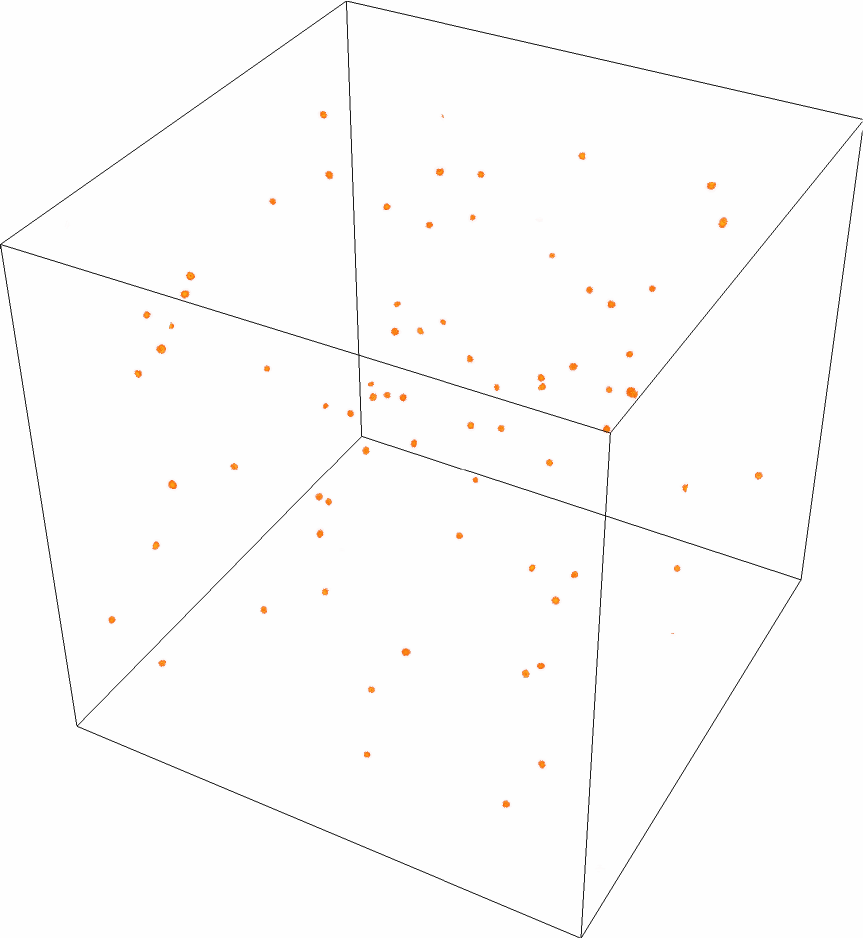}}
\hspace*{\fill}

\caption{Plot of
$|\Tr(\Phi^3)|<0.08$ for $\epsilon=0.006$ for different time steps.}
\label{fig:trphicube}
\end{figure*}

The domain wall network together with the monopoles can be visualized in a 3D plot by looking at the values of $\Tr(\Phi^3)$. At the center of the domain wall $\Tr(\Phi^3)$ flips sign and thus has to vanish at a point. Since the profile functions of the magnetic monopole also flips sign close to the center of the monopole, both defects are visible in a plot of small $|\Tr(\Phi^3)|$. In Fig.~\ref{fig:trphicube}, we plot where the field is $|\Tr(\Phi^3)|<0.08$ for a run with $\epsilon=0.006$ at different time steps $t=200-900$.

At early times, we see snapshots of the domain wall network. In the first time step after the phase transition $t=200$, it looks like one big domain wall. Due to the pressure difference on both sides of the domain walls, the network shrinks and fragments into several collapsing pieces, as seen at time steps $t=300-500$. Starting from time step $t=500$, the network does not look chaotic anymore and there are only a few domain walls left that are shrinking with time. At time step $t=600$, there are two domain walls left (periodic boundary conditions!), one with a spherical shape and one that is stretched in one direction. In the snapshot of $t=700$, the domain walls have collapsed. In all plots, we can see little spherical points, which are magnetic monopoles. 

\begin{figure}
\includegraphics[width=0.45\textwidth]{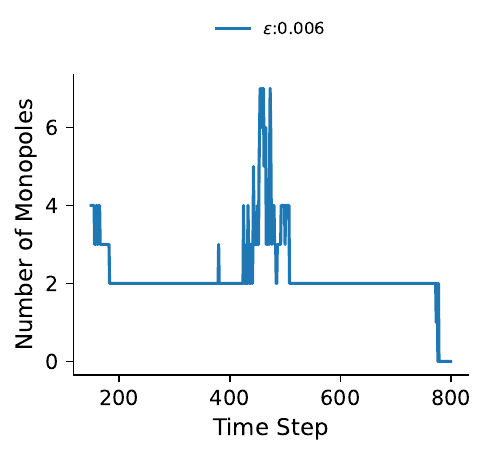}
\caption{Number of monopoles vs. time for a single run
for $\epsilon=0.006$ on a $100^3$ lattice. 
}
\label{spikes}
\end{figure}

There are already some magnetic monopoles present in the first snapshot at $t=200$. However, we see a large amount of monopoles created from collapsing domain walls that carry magnetic charge. These domain walls with magnetic charge can either be created right away from the phase transition, or when domain walls sweep magnetic monopoles. The sweeping mechanism can also cancel magnetic charge, for example when a $+1$ charged domain wall sweeps up an antimonopole with charge $-1$. We also see the creation of monopole-antimonopole pairs when domain walls collapse. 
Especially non-spherical domain walls seem to create several monopole-antimonopole pairs at their cone-like end. Often the creation is momentary and the monopole-antimonpole subsequently annihilate as
can be seen in Fig.~\ref{spikes}. 

When the domain walls are gone at $t=700$ in
Fig.~\ref{fig:trphicube}, the plot shows the magnetic monopoles with better contrast than before. There are two regions where several monopoles and antimonopoles are clustered together. These are are the leftover monopoles from the two collapsed domain walls. At time steps $t=800$ and $t=900$, the plot shows all the monopoles that are left in the simulation. There is some monopole-antimonopole annihilation going on during these time steps. The time evolution of monopoles and antimonopoles has been discussed in~\cite{Zeldovich:1978wj,Preskill:1979zi,Martins:2008zz,Sousa:2017wvx,Hindmarsh:2025vxh}, though under the assumption of a uniform distribution of monopoles whereas the monopoles and antimonopoles produced from the collapse of closed domain walls are clustered.

The animations belonging to the snapshots in Fig.~\ref{fig:potentialenergy} and~\ref{fig:trphicube} can be found in \url{https://youtu.be/EMI41H1adOM}.

\subsection{Number of Monopoles and Domain Wall Area}

\begin{figure}
\includegraphics[width=\linewidth, keepaspectratio]{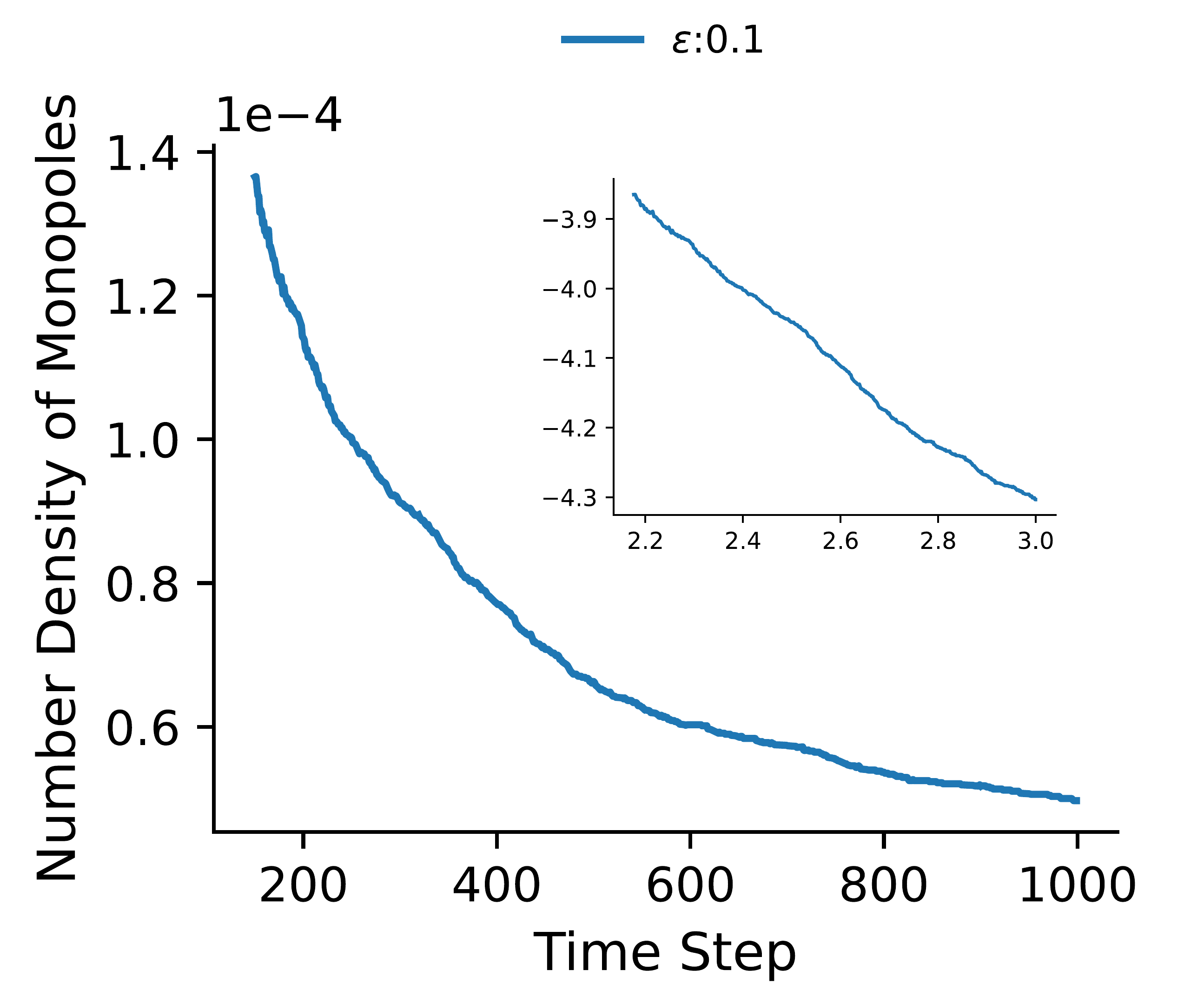}
\caption{Plot of the number density of monopoles vs. time for $\epsilon =0.1$ on a $300^3$ lattice when there are essentially no walls produced. The log-log plot in the inset shows that the number density decays as $\sim t^{-1/2}$.}
\label{monolargeeps}
\end{figure}

\begin{figure*}
\includegraphics[width=0.45\textwidth]{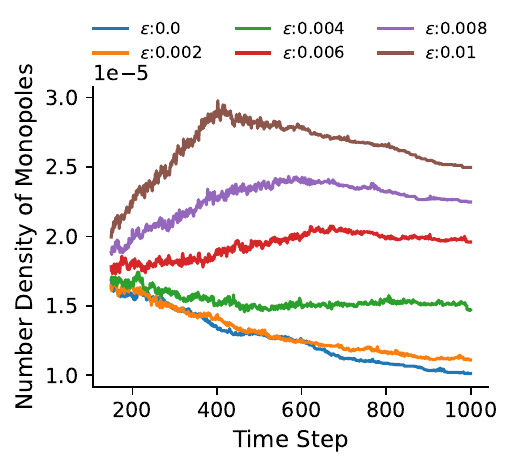} \quad
\includegraphics[width=0.45\textwidth]{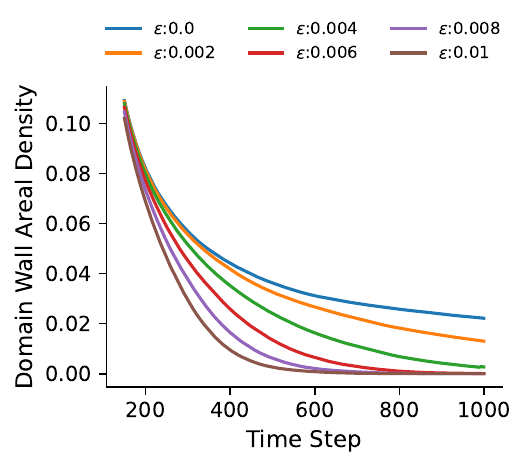}
\caption{Results for our simulation for different values of $\epsilon$, averaging over $10$ runs for each value of $\epsilon$. {\it Left:} Plot of the number density of monopoles. {\it Right:} Plot of density of domain wall area.}
\label{fig:domainwallareovstime}
\end{figure*}

With the monopole detection algorithm described in Sec.~\ref{monoalgo}, we calculate the number of monopoles in our simulation at every time step for different parameters of the potential, for different $\epsilon$. 
First we plot the number density of monopoles in the case with $\epsilon=0.1$ in Fig.~\ref{monolargeeps}. This $\epsilon$ is large enough that essentially no domain walls are produced. We then consider smaller values of $\epsilon$.
On the left side of Fig.~\ref{fig:domainwallareovstime}, we plot the number density of monopoles vs. time for $\epsilon$ ranging from $0$ to $0.01$ and averaging over $10$ runs for each $\epsilon$. For the same runs, we also calculate the domain wall area with the method described previously. The density of domain wall area vs time is plotted on the right side of 
Fig.~\ref{fig:domainwallareovstime} for the same averaged $10$ runs and for the same $\epsilon$ values. 

Both the magnetic monopole detection algorithm and the domain wall area algorithm is not reliable at early times ($t \lesssim 200$) because the field has not settled to its true minimum yet, as can be seen in Fig.~\ref{fig:Phi2_vs_Time}. This leads to some amount of uncertainty at early times, but once the field is close to its VEV, the algorithms work very well. We have also encountered situations in which a collapsing domain wall produced several monopole-antimonopole pairs due to fluctuations of magnetic charge spread over the wall. These monopole-antimonopole pairs are created very close to each other, leading to quick annihilation which can show up as fluctuations in the plot of the number density of monopoles (Fig.~\ref{spikes}).

Most importantly, our results in the monopole density plot show that in the simulations with smaller $\epsilon$, fewer magnetic monopoles survive until the end. This was expected in the original paper~\cite{Dvali:1997sa}, since for smaller $\epsilon$, the domain walls are around longer and have more time to sweep up monopoles. For larger values of the bias parameter $\epsilon$, the domain walls annihilate faster and cannot sweep up as many monopoles. This assumed however, that domain walls and magnetic monopoles are well-separated at production.

Our simulations show that the situation is actually more complicated: Right after the phase transition we find some monopoles, but we see a production of monopoles after the phase transition until a time that is correlated with domain wall collapse. For example, the domain wall network in the $\epsilon=0.01$ runs has collapsed at time step approximately $t=400$. Up to this time, we also see a production of monopoles, that has to come from the collapse of magnetically charged domain walls. Thus, not all monopoles are created during the phase transition, but a significant number of monopoles are produced from domain wall collapse some time after the phase transition, at least for larger values of $\epsilon$. The larger $\epsilon$ is, the more the domain wall network fragments and creates smaller collapsing domain walls, which in turn produce more monopoles.

We believe that a large amount of the magnetic charge produced during the phase transition is spread out on domain walls right away. Thus, the phase transition is not just producing uncharged domain walls and magnetic monopoles but also a lot of domain walls with magnetic charge. The sweeping of monopoles by domain walls is also happening to some extent. For smaller values of $\epsilon$, the domain wall network has more time to interact and in that process the magnetic charge spreads and annihilates on the wall. Thus, there are fewer magnetic monopoles for smaller $\epsilon$.

For $\epsilon=0$, the domain walls are topological and thus stable. The network spreads out over time, reducing the domain wall area density. However, the domain walls do not collapse fully. We see a similar behavior of the domain wall network in the runs with smallest $\epsilon$, {\it e.g.} $\epsilon =0.002$ and to some extent for $\epsilon=0.004$. In those runs, the domain walls also do not fully annihilate until the end of our simulation at time step $t=999$. Since there can be a repulsive force between a domain wall and an anti-domain wall~\cite{Pogosian:2001pq},
there can be a small non-zero $\epsilon$ for which the domain walls, even though they are biased, do not annihilate 
in our simulations. Instead, with periodic boundary conditions, it is possible for the system to form a lattice of domain walls~\cite{Pogosian:2002ua,Antunes:2003be,Antunes:2004ir} though we expect the lattice to eventually decay due to instabilities~\cite{Pogosian:2002ua}.

In Fig.~\ref{fig:domainwallareovstime} we notice that there is an increase in the number of monopoles at early times for larger values of $\epsilon$. We attribute this increase to monopole creation due to collapsing domain walls as the turnover time in those plots roughly coincides with the time at which the domain wall network decays as shown on the right-hand plot in Fig.~\ref{fig:domainwallareovstime}.
For those runs with weakly biased domain walls we do not see a production of monopoles through the collapse of magnetically charged domain walls. We interpret this as a continual sweeping process -- even as small closed domain walls collapse and produce monopoles, the larger domain walls, that are always present in these runs, sweep them up and reduce the number of monopoles.

\subsection{Varying Damping Parameter}
\label{damping}

\begin{figure*}
\includegraphics[width=0.48\textwidth]{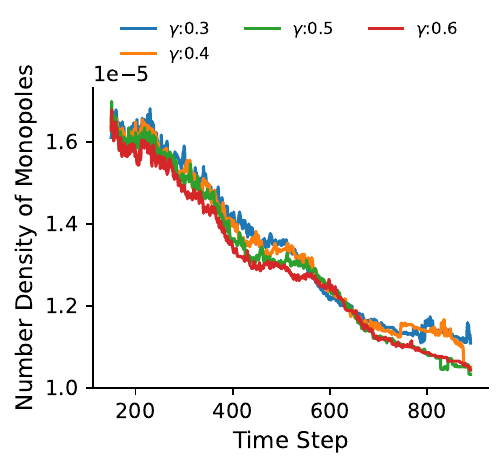} \quad
\includegraphics[width=0.48\textwidth]{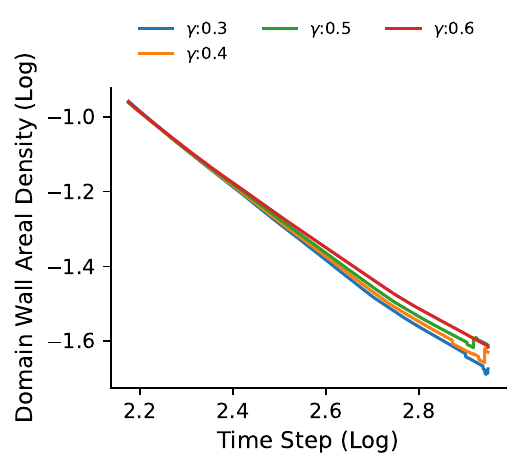}
\caption{Results for our simulation for different values of the damping parameter $\gamma$ and with $\epsilon=0$, averaging over $10$ runs for each value of $\gamma$. {\it Left:} Plot of the number density of monopoles. {\it Right:} Plot of density of domain wall area.}
\label{fig:dampingresults}
\end{figure*}

We ran our simulation with different damping parameters ranging from $\gamma=0.3$ to $\gamma=0.6$, and the results of those runs are shown in Fig.~\ref{fig:dampingresults}. All previous results discussed in this section used a damping parameter of $\gamma=0.6$. With varying the damping parameter, we can check whether our results of monopole number density and domain wall area depend on this parameter. 

We again ran $10$ runs for each value of $\gamma$ and the domain wall area is only slightly lower for the smaller damping parameter. The number of monopoles is only affected at later times, when monopole-antimonopole annihilation becomes important. 
From these plots, we conclude that the damping parameter does not have a significant effect on our results, though it is still possible that more extreme values of the damping parameter may change the outcome quantitatively.

\section{Discussion}
\label{discussion}

In our simulations, we did not incorporate the expansion of the universe. This, however, can have a huge influence on the number of monopoles that survive until today. 
We expect the domain wall network to last longer in an expanding universe because the expansion stretches the domain walls and works against their collapse. Therefore, in an expanding universe, the domain wall network survives longer and has more time to sweep up the monopoles and for the magnetic charge to cancel out. 
We plan to include the expansion of the universe in future simulations.

In Ref.~\cite{Dvali:1997sa} the sweeping mechanism was proposed with estimates of the bias parameter $\epsgut$
that can solve the monopole problem~\footnote{We use $\epsgut$ to denote the bias parameter in a cosmological setting, as opposed to $\epsilon$ in our flat space simulations.}.
The first constraint comes from the requirement that the domain walls not dominate the energy density of the universe~\footnote{This is an overly strong requirement since it is possible to have a brief period of domain wall domination in the early universe and not be in conflict with any cosmological observations. For example, the domain wall network could collapse prior to big bang nucleosynthesis (BBN), the universe could then thermalize and then BBN could follow as in standard cosmology.}.
For the domain walls to never dominate the energy density of the universe, we require $\Delta V > G \sigma^2$~\cite{Vilenkin:2000jqa}, where $\Delta V$ is the energy difference between the two sides of the domain wall due to the bias parameter $\epsgut$, $G = M_P^{-2}$ is Newton's constant and $\sigma$ is the domain wall tension. 
As discussed below \eqref{eq:constants}, $\Delta V \approx \epsgut \etagut^4$ where $\etagut \simeq 10^{14}$ to $10^{16}\;$GeV. The tension of the domain walls is $\sigma \simeq\sqrt{\lambda} \, \etagut^3$ so that we get the bound
\be
\epsgut \gtrsim G \lambda \etagut^2 \approx \lambda \left( \frac{\etagut}{M_P} \right)^2 .
\ee
This bound gives $\epsgut \gtrsim 10^{-10}$ to $10^{-6}$, 
depending on the value of $\lambda$ and $\etagut$. 

Another constraint arises from requiring that the domain wall network survive long enough
to reduce the number density of monopoles to an acceptable level. 
The time at which the domain wall network annihilates is given by balancing the domain wall force of tension, $\sigma/R$, where $R$ is the radius of curvature of the wall, to the force due to the bias, $\Delta V$,
which gives $R\simeq \sqrt{\lambda}/(\epsgut \,\etagut)$. If we assume scaling for the domain wall network and, as above, consider only the case where the cosmology is radiation dominated, then $R \sim t$, where $t$ is the cosmic time. 
Therefore the network annihilates at $t_{\rm ann} \simeq \sqrt{\lambda}/(\epsgut \, \etagut)$~\cite{Hiramatsu:2010yz,Kawasaki:2011vv,Hiramatsu:2013qaa,Ferreira:2022zzo,Kitajima:2023cek,Blasi:2025tmn,Babichev:2025stm,Barbini:2026edx,Gouttenoire:2023ftk,Gouttenoire:2023gbn,Gouttenoire:2025ofv}. 
We expect more efficient monopole annihilation if $t_{\rm ann}$ is large compared to the time of the phase transition, $t_{\rm GUT}$, giving
\be
\epsgut \lesssim \frac{\sqrt{\lambda}}{\etagut \, t_{\rm GUT}} \simeq \sqrt{\lambda} \frac{\etagut}{M_{\rm P}}
\ee
where we used $t_{\rm GUT} = t_{\rm P} (M_{\rm P}/\etagut )^2$.
Altogether the two bounds give a parameter range for which the domain walls can potentially solve the cosmological monopole problem
\be
\lambda \left( \frac{\etagut}{M_P} \right)^2  \lesssim \epsgut \lesssim \sqrt{\lambda} \frac{\etagut}{M_{\rm P}}.
\label{bounds}
\ee
A more detailed discussion may be found in Ref.~\cite{Dvali:1997sa} though we note a few differences. The upper bound in \eqref{bounds} differs from that in Ref.~\cite{Dvali:1997sa} since \cite{Dvali:1997sa} did not require that $t_{\rm ann}$ be much larger than $t_{\rm GUT}$. We impose this requirement as we have assumed that the wall network is scaling which is valid only at times much later than the formation time. A second point to note is that Ref.~\cite{Dvali:1997sa} did not consider the production of monopoles during domain wall collapse. We see evidence of this process in our runs as discussed in Sec.~\ref{results} and this complicates the discussion of whether sweeping can significantly reduce the monopole density to resolve the cosmological monopole problem. We leave it to future work to determine if there is a parameter range for which the sweeping mechanism can definitively solve the monopole problem.




\section{Conclusion}
\label{conclusion}

Symmetry breaking in Grand Unified models, such as minimal $SU(5)$, are based on the vacuum expectation value of an adjoint scalar field, $\Phi$. The adjoint representation is used so that the residual symmetry contains a $U(1)$ factor corresponding to the hypercharge symmetry. Further, the GUT symmetry group does not contain the transformation $\Phi \to -\Phi$ since $\Tr(\Phi^3)$ is invariant under SU(5) 
but odd under $\Phi \to -\Phi$. This means that without a $\Tr(\Phi^3)$ term in the potential, the symmetry of the model is $SU(5)\times Z_2$. The VEV of $\Phi$ will break the $Z_2$ symmetry and produce topological domain walls. Thus GUTs does not only have a cosmological magnetic monopole problem but also a domain wall problem if there is no $\Tr(\Phi^3)$ term in the potential and it is essential to include such a term to be possibly consistent with cosmology. In this work we have examined the consequences of including the $\Tr(\Phi^3)$ term in the potential with coupling strength given by $\epsilon$. This term explicitly breaks the $Z_2$ symmetry.

If $\epsilon$ is large, domain wall formation does not occur but monopole formation proceeds as has been discussed in the literature~\cite{Preskill:1979zi}. 
A more interesting situation occurs if $\epsilon$ is non-zero but small. Then GUT symmetry breaking produces both biased domain walls and magnetic monopoles and these interact strongly with each other.

For $\epsilon =0.1$, there are essentially no domain walls in the system and the monopole number decays as $1/\sqrt{t}$.
For smaller values of $\epsilon$ but still relatively large, magnetic monopoles are predominantly formed by the collapse of closed domain walls until the wall network decays. After that the monopole density slowly decays or even remains frozen. For very small values of $\epsilon$, fewer monopoles are produced and their number continuously decays. In these runs domain walls are always present and we interpret the continuous decay of monopole number as the continuous sweeping of monopoles by the domain walls.

Our simulations show that the sweeping picture is likely too simple since the domain walls can also shed off clustered monopoles and antimonopoles as they collapse. The clustering accelerates monopole annihilation but it is quite possible that some monopoles will survive.

A tantalizing possibility is that closed domain walls can collapse to form Reissner-Nordstrom black holes with magnetic charge, thus sequestering monopoles within black holes~\cite{Vachaspati:2017hjw,Ferrer:2018uiu,Gouttenoire:2023gbn,Hemming:2026osi}. Depending on the frequency of this process, such black holes can contribute to the cosmological dark matter density. Another prediction of our scenario is that the domain walls will emit gravitational waves and generate a stochastic gravitational wave background. Since this process is occurring in the early universe, the gravitational waves will be at a high frequency that depends on the time at which the network collapses. We plan to investigate some of these observable signatures of the GUT epoch in subsequent work.

Our analysis is the first time GUT-like symmetry breaking has been studied numerically but it is certainly not the final word. We plan to include cosmological expansion in the future. This will require larger simulations to offset the limited dynamical range due to an expanding lattice. Then we can directly study the sweeping scenario proposed in Ref.~\cite{Dvali:1997sa}.
The inclusion of weak field gravity can inform the stochastic gravitational wave background. Calculations of the gravitational potential may help estimate the frequency of black hole formation.

\begin{acknowledgments}
We are grateful to Yann Gouttenoire, Fabrizio Rompineve, and Alex Vikman for comments, and to Gil Speyer for assistance with computing. H.H is further grateful to Juanjo Garcia Mesa for assistance with vectorizing the code and achieving significant speed-up using GPU acceleration. Computational resources were provided by the Sol supercomputer 
at Arizona State University \cite{jennewein2023sol}. A.W. would like to thank Maximilian Bachmaier for valuable discussions.
T.V. was supported in part during this project by the U.S. Department of Energy, Office of High Energy 
Physics, under Award No.~DE-SC0019470.
\end{acknowledgments}

\appendix

\bibstyle{aps}
\bibliography{paper}

@article{Hemming:2026osi,
    author = "Hemming, Harish and Vachaspati, Tanmay and Wachowitz, Anja",
    title = "{Domain walls and magnetic monopoles in Grand Unified Models}",
    eprint = "2606.14996",
    archivePrefix = "arXiv",
    primaryClass = "hep-ph",
    month = "6",
    year = "2026",
    journal = ""
}

@article{Zurek:1996sj,
    author = "Zurek, W. H.",
    title = "{Cosmological experiments in condensed matter systems}",
    eprint = "cond-mat/9607135",
    archivePrefix = "arXiv",
    reportNumber = "LA-UR-95-2269, LAUR-95-2269-(LOS-ALAMOS)",
    doi = "10.1016/S0370-1573(96)00009-9",
    journal = "Phys. Rept.",
    volume = "276",
    pages = "177--221",
    year = "1996"
}

@article{Senjanovic:2025enc,
    author = "Senjanovi{\'c}, Goran and Zantedeschi, Michael",
    title = "{Minimal Pati-Salam theory: From cosmic defects to gravitational waves and colliders}",
    eprint = "2504.01893",
    archivePrefix = "arXiv",
    primaryClass = "hep-ph",
    doi = "10.1103/1ld7-mg95",
    journal = "Phys. Rev. D",
    volume = "112",
    number = "5",
    pages = "055018",
    year = "2025"
}

@article{Ferrer:2018uiu,
    author = "Ferrer, Francesc and Masso, Eduard and Panico, Giuliano and Pujolas, Oriol and Rompineve, Fabrizio",
    title = "{Primordial Black Holes from the QCD axion}",
    eprint = "1807.01707",
    archivePrefix = "arXiv",
    primaryClass = "hep-ph",
    doi = "10.1103/PhysRevLett.122.101301",
    journal = "Phys. Rev. Lett.",
    volume = "122",
    number = "10",
    pages = "101301",
    year = "2019"
}

@article{Ferreira:2022zzo,
    author = "Ferreira, Ricardo Z. and Notari, Alessio and Pujolas, Oriol and Rompineve, Fabrizio",
    title = "{Gravitational waves from domain walls in Pulsar Timing Array datasets}",
    eprint = "2204.04228",
    archivePrefix = "arXiv",
    primaryClass = "astro-ph.CO",
    reportNumber = "CERN-TH-2022-214",
    doi = "10.1088/1475-7516/2023/02/001",
    journal = "JCAP",
    volume = "02",
    pages = "001",
    year = "2023"
}

@article{Vachaspati:2017hjw,
    author = "Vachaspati, Tanmay",
    title = "{Lunar Mass Black Holes from QCD Axion Cosmology}",
    eprint = "1706.03868",
    archivePrefix = "arXiv",
    primaryClass = "hep-th",
    month = "6",
    year = "2017",
    journal = ""
}

@article{Hiramatsu:2013qaa,
    author = "Hiramatsu, Takashi and Kawasaki, Masahiro and Saikawa, Ken'ichi",
    title = "{On the estimation of gravitational wave spectrum from cosmic domain walls}",
    eprint = "1309.5001",
    archivePrefix = "arXiv",
    primaryClass = "astro-ph.CO",
    reportNumber = "ICRR-REPORT-659-2013-8, IPMU13-0182, YITP-13-87",
    doi = "10.1088/1475-7516/2014/02/031",
    journal = "JCAP",
    volume = "02",
    pages = "031",
    year = "2014"
}

@article{Kawasaki:2011vv,
    author = "Kawasaki, Masahiro and Saikawa, Ken'ichi",
    title = "{Study of gravitational radiation from cosmic domain walls}",
    eprint = "1102.5628",
    archivePrefix = "arXiv",
    primaryClass = "astro-ph.CO",
    reportNumber = "ICRR-REPORT-581-2010-14, IPMU11-0032",
    doi = "10.1088/1475-7516/2011/09/008",
    journal = "JCAP",
    volume = "09",
    pages = "008",
    year = "2011"
}

@article{Hiramatsu:2010yz,
    author = "Hiramatsu, Takashi and Kawasaki, Masahiro and Saikawa, Ken'ichi",
    title = "{Gravitational Waves from Collapsing Domain Walls}",
    eprint = "1002.1555",
    archivePrefix = "arXiv",
    primaryClass = "astro-ph.CO",
    reportNumber = "ICRR-REPORT-559-2009-21, IPMU10-0024",
    doi = "10.1088/1475-7516/2010/05/032",
    journal = "JCAP",
    volume = "05",
    pages = "032",
    year = "2010"
}

@article{Blasi:2025tmn,
    author = {Blasi, Simone and Mariotti, Alberto and Rase, A{\"a}ron and Vanvlasselaer, Miguel},
    title = "{Domain walls in the scaling regime: Equal Time Correlator and Gravitational Waves}",
    eprint = "2511.16649",
    archivePrefix = "arXiv",
    primaryClass = "hep-ph",
    month = "11",
    year = "2025",
    journal = ""
}

@article{Kitajima:2023cek,
    author = "Kitajima, Naoya and Lee, Junseok and Murai, Kai and Takahashi, Fuminobu and Yin, Wen",
    title = "{Gravitational waves from domain wall collapse, and application to nanohertz signals with QCD-coupled axions}",
    eprint = "2306.17146",
    archivePrefix = "arXiv",
    primaryClass = "hep-ph",
    reportNumber = "TU-1198",
    doi = "10.1016/j.physletb.2024.138586",
    journal = "Phys. Lett. B",
    volume = "851",
    pages = "138586",
    year = "2024"
}

@article{Leese:1990cj,
    author = "Leese, Robert and Prokopec, Tomislav",
    title = "{Clustering of cosmological defects at the time of formation}",
    reportNumber = "BROWN-HET-778",
    doi = "10.1016/0370-2693(91)90964-R",
    journal = "Phys. Lett. B",
    volume = "260",
    pages = "27--31",
    year = "1991"
}

@article{Sousa:2017wvx,
    author = "Sousa, Lara and Avelino, Pedro P.",
    title = "{Revisiting the velocity-dependent one-scale model for monopoles}",
    eprint = "1703.09054",
    archivePrefix = "arXiv",
    primaryClass = "astro-ph.CO",
    doi = "10.1103/PhysRevD.96.023521",
    journal = "Phys. Rev. D",
    volume = "96",
    number = "2",
    pages = "023521",
    year = "2017"
}

@article{Martins:2008zz,
    author = "Martins, C. J. A. P. and Achucarro, A.",
    title = "{Evolution of local and global monopole networks}",
    doi = "10.1103/PhysRevD.78.083541",
    journal = "Phys. Rev. D",
    volume = "78",
    pages = "083541",
    year = "2008"
}

@article{Hindmarsh:2025vxh,
    author = {Hindmarsh, Mark and Lopez-Eiguren, Asier and Sepp{\"a}, Riikka and Weir, David J.},
    title = "{Numerical simulations of magnetic monopole evolution in an expanding universe}",
    eprint = "2511.14204",
    archivePrefix = "arXiv",
    primaryClass = "astro-ph.CO",
    reportNumber = "HIP-2025-31/TH",
    month = "11",
    year = "2025",
    journal = ""
}

@article{Preskill:1979zi,
    author = "Preskill, John",
    title = "{Cosmological Production of Superheavy Magnetic Monopoles}",
    reportNumber = "HUTP-79/A028",
    doi = "10.1103/PhysRevLett.43.1365",
    journal = "Phys. Rev. Lett.",
    volume = "43",
    pages = "1365",
    year = "1979"
}

@article{Zeldovich:1978wj,
    author = "Zeldovich, Ya. B. and Khlopov, M. Yu.",
    title = "{On the Concentration of Relic Magnetic Monopoles in the Universe}",
    doi = "10.1016/0370-2693(78)90232-0",
    journal = "Phys. Lett. B",
    volume = "79",
    pages = "239--241",
    year = "1978"
}

@article{Vachaspati:2016abz,
    author = "Vachaspati, Tanmay",
    title = "{Creation of Magnetic Monopoles in Classical Scattering}",
    eprint = "1607.07460",
    archivePrefix = "arXiv",
    primaryClass = "hep-th",
    doi = "10.1103/PhysRevLett.117.181601",
    journal = "Phys. Rev. Lett.",
    volume = "117",
    number = "18",
    pages = "181601",
    year = "2016"
}

@article{Wilkinson:1978zh,
    author = "Wilkinson, David and Bais, F. Alexander",
    title = "{Exact SU(N) Monopole Solutions with Spherical Symmetry}",
    reportNumber = "FERMILAB-PUB-78-077-T",
    doi = "10.1103/PhysRevD.19.2410",
    journal = "Phys. Rev. D",
    volume = "19",
    pages = "2410",
    year = "1979"
}

@article{Bais:1978yh,
    author = "Bais, F. Alexander and Weldon, H. Arthur",
    title = "{Exact Monopole Solutions in SU($n$) Gauge Theory}",
    reportNumber = "Print-78-0709 (PENN)",
    doi = "10.1103/PhysRevLett.41.601",
    journal = "Phys. Rev. Lett.",
    volume = "41",
    pages = "601",
    year = "1978"
}

@article{Bachmaier:2023zmq,
    author = "Bachmaier, Maximilian and Dvali, Gia and Valbuena-Berm{\'u}dez, Juan Sebasti{\'a}n",
    title = "{Radiation emission during the erasure of magnetic monopoles}",
    eprint = "2306.12958",
    archivePrefix = "arXiv",
    primaryClass = "hep-th",
    doi = "10.1103/PhysRevD.108.103501",
    journal = "Phys. Rev. D",
    volume = "108",
    number = "10",
    pages = "103501",
    year = "2023"
}

@article{Dvali:2022rgx,
    author = "Dvali, Gia and Valbuena-Berm{\'u}dez, Juan Sebasti{\'a}n",
    title = "{Erasure of strings and vortices}",
    eprint = "2212.07535",
    archivePrefix = "arXiv",
    primaryClass = "hep-th",
    doi = "10.1103/PhysRevD.107.035001",
    journal = "Phys. Rev. D",
    volume = "107",
    number = "3",
    pages = "035001",
    year = "2023"
}

@article{Vachaspati:2003zp,
    author = "Vachaspati, Tanmay",
    title = "{Symmetries within domain walls}",
    eprint = "hep-th/0303137",
    archivePrefix = "arXiv",
    doi = "10.1103/PhysRevD.67.125002",
    journal = "Phys. Rev. D",
    volume = "67",
    pages = "125002",
    year = "2003"
}

@article{Vachaspati:2001pw,
    author = "Vachaspati, Tanmay",
    title = "{A Class of kinks in SU(N) x Z(2)}",
    eprint = "hep-th/0102047",
    archivePrefix = "arXiv",
    doi = "10.1103/PhysRevD.63.105010",
    journal = "Phys. Rev. D",
    volume = "63",
    pages = "105010",
    year = "2001"
}

@article{Pogosian:1999zi,
    author = "Pogosian, Levon and Vachaspati, Tanmay",
    title = "{Interaction of magnetic monopoles and domain walls}",
    eprint = "hep-ph/9909543",
    archivePrefix = "arXiv",
    reportNumber = "CWRU-P32-99",
    doi = "10.1103/PhysRevD.62.105005",
    journal = "Phys. Rev. D",
    volume = "62",
    pages = "105005",
    year = "2000"
}

@article{Pogosian:2000xv,
    author = "Pogosian, Levon and Vachaspati, Tanmay",
    title = "{Domain walls in SU(5)}",
    eprint = "hep-ph/0007045",
    archivePrefix = "arXiv",
    reportNumber = "CWRU-P6-00",
    doi = "10.1103/PhysRevD.62.123506",
    journal = "Phys. Rev. D",
    volume = "62",
    pages = "123506",
    year = "2000"
}

@article{Pogosian:2001fm,
    author = "Pogosian, Levon and Vachaspati, Tanmay",
    title = "{Space of kink solutions in SU(N) * Z(2)}",
    eprint = "hep-th/0105128",
    archivePrefix = "arXiv",
    doi = "10.1103/PhysRevD.64.105023",
    journal = "Phys. Rev. D",
    volume = "64",
    pages = "105023",
    year = "2001"
}

@article{Pogosian:2002ua,
    author = "Pogosian, Levon and Vachaspati, Tanmay",
    title = "{Domain wall lattices}",
    eprint = "hep-th/0210232",
    archivePrefix = "arXiv",
    doi = "10.1103/PhysRevD.67.065012",
    journal = "Phys. Rev. D",
    volume = "67",
    pages = "065012",
    year = "2003"
}

@article{Antunes:2003be,
    author = "Antunes, Nuno D. and Pogosian, Levon and Vachaspati, Tanmay",
    title = "{On formation of domain wall lattices}",
    eprint = "hep-ph/0307349",
    archivePrefix = "arXiv",
    doi = "10.1103/PhysRevD.69.043513",
    journal = "Phys. Rev. D",
    volume = "69",
    pages = "043513",
    year = "2004"
}

@article{Brush:2015vda,
    author = "Brush, Micah and Pogosian, Levon and Vachaspati, Tanmay",
    title = "{Magnetic monopole{\textemdash}domain wall collisions}",
    eprint = "1505.08170",
    archivePrefix = "arXiv",
    primaryClass = "hep-th",
    doi = "10.1103/PhysRevD.92.045008",
    journal = "Phys. Rev. D",
    volume = "92",
    number = "4",
    pages = "045008",
    year = "2015"
}

@book{Vachaspati:2006zz,
    author = "Vachaspati, Tanmay",
    title = "{Kinks and Domain Walls : An Introduction to Classical and Quantum Solitons}",
    doi = "10.1017/9781009290456",
    isbn = "978-1-009-29045-6, 978-1-009-29041-8, 978-1-009-29042-5, 978-0-521-14191-8, 978-0-521-83605-0, 978-0-511-24290-8",
    publisher = "Oxford University Press",
    year = "2007"
}

@article{Dvali:1997sa,
    author = "Dvali, G. R. and Liu, Hong and Vachaspati, Tanmay",
    title = "{Sweeping away the monopole problem}",
    eprint = "hep-ph/9710301",
    archivePrefix = "arXiv",
    reportNumber = "CWRU-P15-97, IMPERIAL-TP-97-98-3, CERN-TH-97-273",
    doi = "10.1103/PhysRevLett.80.2281",
    journal = "Phys. Rev. Lett.",
    volume = "80",
    pages = "2281--2284",
    year = "1998"
}

@article{Ng:2008mp,
    author = "Ng, Yifung and Kibble, T. W. B. and Vachaspati, Tanmay",
    title = "{Formation of Non-Abelian Monopoles Connected by Strings}",
    eprint = "0806.0155",
    archivePrefix = "arXiv",
    primaryClass = "hep-th",
    doi = "10.1103/PhysRevD.78.046001",
    journal = "Phys. Rev. D",
    volume = "78",
    pages = "046001",
    year = "2008"
}

@book{Vilenkin:2000jqa,
    author = "Vilenkin, A. and Shellard, E. P. S.",
    title = "{Cosmic Strings and Other Topological Defects}",
    isbn = "978-0-521-65476-0",
    publisher = "Cambridge University Press",
    month = "7",
    year = "2000"
}

@article{Kibble:1976sj,
    author = "Kibble, T. W. B.",
    title = "{Topology of Cosmic Domains and Strings}",
    reportNumber = "ICTP/75/5",
    doi = "10.1088/0305-4470/9/8/029",
    journal = "J. Phys. A",
    volume = "9",
    pages = "1387--1398",
    year = "1976"
}

@article{Zhang:2019vsb,
    author = "Zhang, Yiyang and Vachaspati, Tanmay and Ferrer, Francesc",
    title = "{Magnetic field production at a first-order electroweak phase transition}",
    eprint = "1902.02751",
    archivePrefix = "arXiv",
    primaryClass = "hep-ph",
    doi = "10.1103/PhysRevD.100.083006",
    journal = "Phys. Rev. D",
    volume = "100",
    number = "8",
    pages = "083006",
    year = "2019"
}

@phdthesis{Bachmaier:2026,
    author = "Bachmaier, Maximilian",
    title = "{The Fate of Magnetic Monopoles in the Early Universe}",
    school = "Munich U.",
    year = "2026"


}

@inproceedings{jennewein2023sol,
  author    = {Jennewein, Douglas M. and Lee, Jay and Kurtz, Christopher and others},
  title     = {The Sol Supercomputer at Arizona State University},
  booktitle = {Practice and Experience in Advanced Research Computing},
  year      = {2023},
  pages     = {296--301},
  publisher = {Association for Computing Machinery},
  doi       = {10.1145/3569951.3597573}
}

@article{Gouttenoire:2023ftk,
    author = "Gouttenoire, Yann and Vitagliano, Edoardo",
    title = "{Domain wall interpretation of the PTA signal confronting black hole overproduction}",
    eprint = "2306.17841",
    archivePrefix = "arXiv",
    primaryClass = "gr-qc",
    doi = "10.1103/PhysRevD.110.L061306",
    journal = "Phys. Rev. D",
    volume = "110",
    number = "6",
    pages = "L061306",
    year = "2024"
}

@article{Gouttenoire:2023gbn,
    author = "Gouttenoire, Yann and Vitagliano, Edoardo",
    title = "{Primordial black holes and wormholes from domain wall networks}",
    eprint = "2311.07670",
    archivePrefix = "arXiv",
    primaryClass = "hep-ph",
    doi = "10.1103/PhysRevD.109.123507",
    journal = "Phys. Rev. D",
    volume = "109",
    number = "12",
    pages = "123507",
    year = "2024"
}

@article{Antunes:2004ir,
    author = "Antunes, Nuno D. and Vachaspati, Tanmay",
    title = "{Spontaneous formation of domain wall lattices in two spatial dimensions}",
    eprint = "hep-ph/0404227",
    archivePrefix = "arXiv",
    doi = "10.1103/PhysRevD.70.063516",
    journal = "Phys. Rev. D",
    volume = "70",
    pages = "063516",
    year = "2004"
}

@article{Gouttenoire:2025ofv,
    author = "Gouttenoire, Yann and King, Stephen F. and Roshan, Rishav and Wang, Xin and White, Graham and Yamazaki, Masahito",
    title = "{Cosmological consequences of domain walls biased by quantum gravity}",
    eprint = "2501.16414",
    archivePrefix = "arXiv",
    primaryClass = "hep-ph",
    doi = "10.1103/7zmx-v16z",
    journal = "Phys. Rev. D",
    volume = "112",
    number = "7",
    pages = "075007",
    year = "2025"
}

@article{Langacker:1980kd,
    author = "Langacker, Paul and Pi, So-Young",
    title = "{Magnetic Monopoles in Grand Unified Theories}",
    reportNumber = "SLAC-PUB-2496, Print-80-0262 (IAS,PRINCETON)",
    doi = "10.1103/PhysRevLett.45.1",
    journal = "Phys. Rev. Lett.",
    volume = "45",
    pages = "1",
    year = "1980"
}

@article{Pogosian:2001pq,
    author = "Pogosian, Levon",
    title = "{Kink interactions in SU(N) x Z(2)}",
    eprint = "hep-th/0111206",
    archivePrefix = "arXiv",
    doi = "10.1103/PhysRevD.65.065023",
    journal = "Phys. Rev. D",
    volume = "65",
    pages = "065023",
    year = "2002"
}

@article{Babichev:2025stm,
    author = "Babichev, E. and Dankovsky, I. and Gorbunov, D. and Ramazanov, S. and Vikman, A.",
    title = "{Biased domain walls: faster annihilation, weaker gravitational waves}",
    eprint = "2504.07902",
    archivePrefix = "arXiv",
    primaryClass = "hep-ph",
    doi = "10.1088/1475-7516/2025/10/103",
    journal = "JCAP",
    volume = "10",
    pages = "103",
    year = "2025"
}

@article{Barbini:2026edx,
    author = "Barbini, Davide and Notari, Alessio and Pujol{\`a}s, Oriol and Rompineve, Fabrizio and Torrent{\'\i}, Francisco",
    title = "{Biased Domain Wall Networks and their Gravitational Waves}",
    eprint = "2607.18107",
    archivePrefix = "arXiv",
    primaryClass = "astro-ph.CO",
    month = "7",
    year = "2026",
    journal = ""
}

\end{document}